\PassOptionsToPackage{dvipsnames}{xcolor}
\documentclass[reprint,aps,pra,twocolumn,superscriptaddress,nofootinbib]{revtex4-2}
\pdfoutput=1
\usepackage{lipsum} 
\usepackage{amsthm}
\usepackage{amssymb}
\usepackage{graphicx}
\usepackage{tikz}
\usepackage{array}[=2016-10-06]
\usepackage{makecell}
\usepackage{tabularx}
\usepackage{etoolbox}

\AtBeginEnvironment{thebibliography}{%
    \interlinepenalty=10000
}

\newcolumntype{Y}{>{\centering\arraybackslash}X}

\usepackage[utf8]{inputenc}
\usepackage{amsfonts}

\usepackage{footmisc}
\usepackage{bbold}

\usepackage{mathrsfs}
\usepackage{verbatim}
\usepackage{cancel}
\usepackage[caption=false]{subfig}

\usepackage{xcolor}
\usepackage{ragged2e}
\usepackage{amsmath}

\usepackage{bm}
\usepackage{hyperref}

\newcommand{\ad}{\hat{a}^{\dagger}}

\newcommand{\ket}[1]{\left| {#1}\right\rangle}
\newcommand{\bra}[1]{\left\langle {#1}\right|}
\newcommand{\braket}[2]{\left\langle {#1}\middle|{#2} \right\rangle}

\newcommand{\perm}[1]{\text{perm}\left({#1}\right)}

\DeclareMathOperator{\Perm}{perm}

\newcommand{\facetheadercell}[1]{%
\parbox[c][0.9cm][c]{\linewidth}{\centering #1}%
}

\newcommand{\facetbodycell}[1]{%
\parbox[c][1.7cm][c]{\linewidth}{\centering #1}%
}

\newtheorem{conjecturep}{Conjecture~P\ignorespaces}

\newtheorem{conjecturem}{Conjecture~M\ignorespaces}

\newtheorem{proposition}{Proposition}

\begin{document}
\title{A unified framework for anomalous boson bunching}


\author{Léo Pioge}
\affiliation{Centre for Quantum Information and Communication, \'Ecole polytechnique de Bruxelles, CP 165/59, Universit\'e libre de Bruxelles, 1050 Brussels, Belgium}

\author{Leonardo Novo}
\affiliation{International Iberian Nanotechnology Laboratory (INL), Av. Mestre José Veiga, 4715-330 Braga, Portugal}

\author{Nicolas J. Cerf}
\affiliation{Centre for Quantum Information and Communication, \'Ecole polytechnique de Bruxelles, CP 165/59, Universit\'e libre de Bruxelles, 1050 Brussels, Belgium}

\begin{abstract}
Anomalous bunching is a paradoxical quantum interferometric phenomenon in which partially distinguishable photons exhibit a higher probability of bunching into two or more modes than fully indistinguishable photons~[Nat. Photonics \textbf{17}, 702 (2023)]. While this effect is directly linked to violations of certain conjectures on matrix permanents, the mechanism underlying the anomaly is not yet fully understood. Here, we show that boson bunching is governed not by internal indistinguishability alone, but by the total indistinguishability of the postselected output state. Total indistinguishability combines the internal degrees of freedom, such as polarization or arrival time, with the spatial degrees of freedom of their wave functions restricted to the measured output modes of the interferometer. This reformulation preserves the expected connection between boson bunching and total indistinguishability, while clarifying how anomalous bunching can arise. It also reveals several additional facets of the same phenomenon, which are captured within a unified framework. In particular, we exhibit situations in which adding an independent source of distinguishability, either internal or spatial, can enhance multimode or even single-mode bunching probabilities, offering new insights into the subtle role of distinguishability in multiphoton interference.

\end{abstract}

\maketitle
\vspace{-5pt}

\section{Introduction}
\label{sec:Introduction}

One of the most intriguing consequences of bosonic interference is the tendency of indistinguishable bosons to bunch, i.e., their inclination to occupy the same mode. The Hong--Ou--Mandel (HOM) effect is one of the most widely recognized demonstrations of this phenomenon~\cite{HOM}. In the HOM experiment, two indistinguishable photons enter different input arms of a 50:50 beam splitter. Due to destructive interference between the paths in which both photons are transmitted or both photons are reflected, the photons always exit the interferometer in the same output mode. However, if a distinguishing degree of freedom is introduced—such as a time delay, polarization, or frequency—this destructive interference is weakened, reducing the probability of bunching \cite{HOM}. The recent development of photonic quantum technologies~\cite{pelucchi2022potential} has stimulated intensive studies of both partial distinguishability~\cite{PhysRevLett.132.050201,renema2019classical,jones2020multiparticle,MenssenDistinguishability,shchesnovich2015partial,shchesnovich2015tight,tichy2015_partial_distinguishability} and boson bunching~\cite{tichy2015_partial_distinguishability,rodari2024,Spagnolo_General_Rules,Niu:09,geller2025} in increasingly complex interferometric configurations involving larger numbers of photons and modes.

In this context, the concept of multimode boson bunching was introduced in Ref.~\cite{shchesnovich2016universality}, referring to output events in which all photons are detected within some subset of spatial output modes of a linear interferometer. This notion naturally extends the usual single-mode picture, where all photons are measured in a single spatial mode, with the added benefit that multimode bunching probabilities are larger and hence potentially easier to measure in a large-scale interferometric setup. Turning to multimode bunching was therefore anticipated to yield an effective method to validate the correct functioning of a boson sampler \cite{aaronson2011computational},  providing an experimentally accessible way to distinguish an ideal boson sampler from one with partially distinguishable photons \cite{shchesnovich2016universality, anguita2025experimentalvalidationbosonsampling,Seron2024efficientvalidation,bressanini2024gaussianbosonsamplingvalidation,young2023atomic,pioge2026validation}. In this regard, Refs.~\cite{shchesnovich2016universality,shchesnovich2016permanent} provided the first compelling evidence supporting the validity of what was later coined the ``Multimode Bunching Conjecture'' \cite{bosonbunching}. This conjecture generalizes the HOM effect to the multimode case by asserting that, for any input state of classically correlated bosons, the probability of multimode bunching at the output of any linear interferometer is maximal when the bosons are fully indistinguishable in their internal degrees of freedom (i.e., perfect mode matching). This statement extends the well-established case of single-mode bunching~\cite{tichy2015_partial_distinguishability} and follows from a simple physical intuition: since boson bunching is a manifestation of quantum interference, which is strongest when the particles are indistinguishable, the bunching probability should naturally be maximized in this regime. Several additional considerations further supported this conjecture. Notably, in Ref.~\cite{shchesnovich2016universality}, it was shown to be equivalent to the Bapat--Sunder conjecture \cite{bapat1985majorization} related to matrix permanents. This equivalence implies that the Multimode Bunching Conjecture holds in the one-, two-, and three-photon regimes~\cite{bapat1986extremal}. Furthermore, its counterpart for fermions, which follows from the Oppenheim inequality for determinants~\cite{oppenheim1930inequalities}, can actually be proven \cite{shchesnovich2016universality}.

For all these reasons, it came as a surprise that a peculiar seven-photon interferometric setup can be found in which independent photons with different polarization exhibit stronger bunching in two modes than seven photons with identical polarization, thereby violating the Multimode Bunching Conjecture~\cite{bosonbunching}. This finding was based on a counterexample to the Bapat--Sunder conjecture due to Drury \cite{drury2016}, involving the permanent of a seven-dimensional matrix. The resulting phenomenon, termed \textit{anomalous boson bunching} in Ref.~\cite{pioge2023anomalous}, not only has fundamental significance but also bears practical implications for boson-sampling validation, since it implies that reaching the highest probability of multimode bunching is not necessarily a signature of fully indistinguishable bosons.  Following this discovery, other setups exhibiting anomalous bunching were found, including scenarios involving nearly indistinguishable photons~\cite{pioge2023anomalous} and cases where time delays constitute the sole source of distinguishability~\cite{pioge2026validation}. At the same time, other investigations sought to identify conditions on the interferometer and input state under which anomalous bunching cannot arise~\cite{geller2025,pioge2026validation}, restoring the validity of the Multimode Bunching Conjecture in some restricted regimes.

Altogether, these observations suggest that the role of particle distinguishability in multimode boson bunching remains incompletely understood. In the present work, we address this issue by arguing that the relevant way to characterize indistinguishability in a multimode bunching event should encompass not only the internal degrees of freedom of the wave functions of the input photons, but also their spatial degrees of freedom after restriction to the postselected output modes.  Hence, we introduce the notion of \emph{total indistinguishability} in Sec.~\ref{sec:New_interpretation_of_the_bunching_probability}, whose maximization leads, this time, to the highest multimode bunching probability. Crucially, the maximization of total indistinguishability depends on the chosen interferometer and subset of output modes in such a way that it does not necessarily yield identical photons in their internal degrees of freedom, thereby opening the door to anomalous bunching effects.

Building on this interpretation, we revisit the original setting of anomalous bunching in Sec.~\ref{sec:standard_anomalous_bunching}, exploiting the total indistinguishability. In addition, we uncover three complementary facets of the anomalous bunching phenomenon, which all arise from a violation of the same mathematical inequality but have  different physical meanings. 
The first, discussed in Sec.~\ref{sec:Extra_distinguishability}, appears when considering the probability of all photons bunching into a single spatial mode, when the input photons carry two independent internal degrees of freedom, such as polarization and time.  In particular, we show that, for any interferometer, adding specific time delays between photons can enhance bunching if the photons occupy some particular polarization states. 

The second facet, presented in Sec.~\ref{sec:Multimode_outperform_single-mode}, focuses on the question of how much  multimode bunching probabilities are enhanced due to quantum interference when compared to classical processes. We show that in some scenarios connected to the violation of the Bapat--Sunder conjecture, multimode bunching probabilities undergo a larger quantum enhancement than single-mode bunching probabilities.

The final facet, presented in Sec.~\ref{sec:several_interferometers}, is a purely spatial manifestation of anomalous bunching. In this scenario, we consider several copies of the same interferometer $U$, with the input photons sent in a superposition of spatial modes over these different copies. In this scenario, a violation of the Bapat--Sunder inequality implies that multimode bunching is not always maximized when all the photons are sent to the same interferometer.

\section{MULTIMODE BOSON BUNCHING}
\label{sec:Linear_interferometry}
\subsection{Preliminaries}

\begin{figure}[t]
    \centering
    \includegraphics[width = 0.49\textwidth]{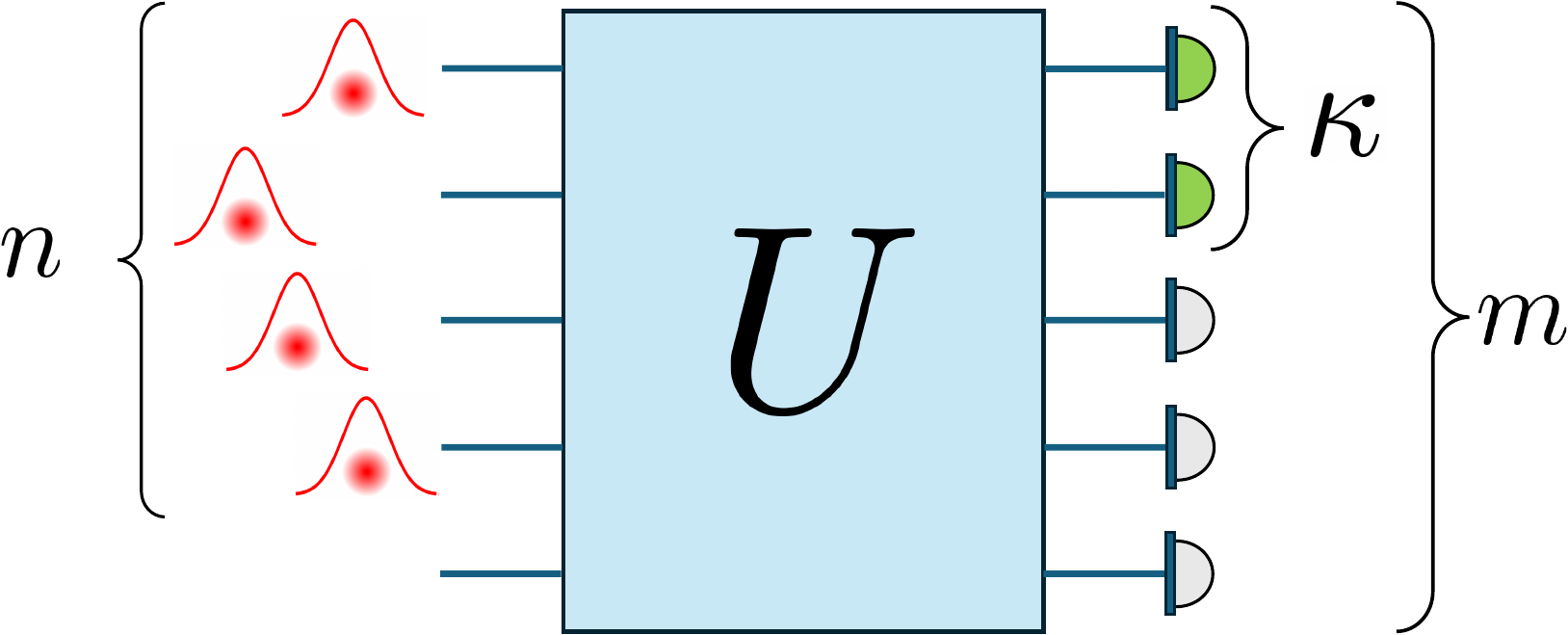}
    \caption{\justifying Schematic representation of an interferometric configuration in which $n$ single photons are injected into the first $n$ spatial modes of an $m$-mode linear interferometer $U$. The time delay between the photons (viewed as an internal degree of freedom) is one possible source of distinguishability. After propagating through the interferometer, the detectors measure the spatial modes occupied by the photons regardless of their time of arrival (which is not resolved). Our analysis focuses on the probability $P_{\kappa}$ that all $n$ photons are detected in a subset $\kappa$ of the output modes (see green detectors).} 
\label{fig:generalInterferometer}
\end{figure}

This section introduces the theoretical framework for multimode boson bunching with partially distinguishable bosons. We consider an $m$-mode linear interferometer, described by a unitary operator $\hat{U}$, in which $n$ single bosons are injected into the first $n$ input modes (see Fig.~\ref{fig:generalInterferometer}). Each photon carries two types of degrees of freedom: a spatial degree of freedom, associated with the spatial mode in the interferometer, and internal degrees of freedom, which comprise all other photonic variables, such as polarization, time delay, frequency, or orbital angular momentum. Throughout this work, we restrict our attention to input configurations prepared in pure, separable states, with the single-photon Hilbert space factorized as $\mathcal{H}=\mathcal{H}_{s} \otimes \mathcal{H}_{int}$, where $\mathcal{H}_{s}$ denotes the spatial Hilbert space and $\mathcal{H}_{int}$ the internal Hilbert space. The input state can then be written as
\begin{equation}
\ket{\Psi^{\mathrm{in}}}=\prod_{i=1}^{n}a^{\dagger}_{i,\phi_i}\ket{0},
\end{equation}
where the creation operator $a^\dagger_{i,\phi_i}$ defines the single-photon state
\begin{equation}
\ket{\psi_i}=a^\dagger_{i,\phi_i}\ket{0}=\ket{i}\otimes\ket{\phi_i}.
\end{equation}
Here, $\ket{i} \in \mathcal{H}_{s}$ denotes the index of the spatial mode occupied by the photon,  while $\ket{\phi_i}\in\mathcal{H}_{int}$ denotes its internal state. The linear interferometer acts only on the spatial degree of freedom, according to
\begin{equation}
\hat{U}a^\dagger_{i,\phi_i}\, \hat{U}^\dagger
=
\sum_{k=1}^m U_{k,i}a^\dagger_{k,\phi_i},
\qquad \forall i.
\end{equation}
At the interferometer's output, the detectors resolve the photons' spatial modes, while the internal degrees of freedom are not measured.

The central quantity investigated in this work is the multimode bunching probability $P_\kappa$, defined as the probability that all photons are detected within the subset $\kappa$ of the output modes. Equivalently, $P_\kappa$ also corresponds to the postselection probability of projecting  all output modes outside of $\kappa$ onto the vacuum state. According to Ref.~\cite{shchesnovich2016universality}, the multimode bunching probability can be expressed as 
\begin{equation}
\label{eq:multimode_boson_bunching}
    P_\kappa= \perm{H\odot S^{\mathrm{int}}}.
\end{equation}
Here, $H$ is an $n \times n$ positive
semidefinite Hermitian ($p.s.d.h.$) matrix such that 
\begin{equation}
\label{eq:definition-matrix-H}
H_{i,j}=\sum_{k\in\kappa}U^{*}_{k,i}U_{k,j},~~\forall i,j.
\end{equation}
Thus, the dependence of $P_\kappa$ on the interferometer $U$ and  measured subset of modes $\kappa$ is entirely contained in $H$. The matrix $S^{\mathrm{int}}$ is the so-called \textit{distinguishability matrix}, and it encapsulates the dependence on photon distinguishability through their pairwise overlaps \cite{tichy2015_partial_distinguishability},
$\displaystyle S^{\mathrm{int}}_{i,j}=\braket{\phi_i}{\phi_j},~~\forall i,j.$
Note that $S^{\mathrm{int}}$ is an $n \times n$ Gram matrix, as it is $p.s.d.h.$ and has ones on its diagonal. Finally, $\perm{\cdot}$ designates the matrix permanent and $\odot$ denotes the Hadamard (element-wise) product, defined explicitly by $(H\odot S^{\mathrm{int}})_{i,j}=H_{i,j}S^{\mathrm{int}}_{i,j}$. Appendix~\ref{appendix:second_quantizationcalculation} presents an alternative derivation of Eq.~\eqref{eq:multimode_boson_bunching} based on the second-quantization formalism and  commutation relations.

Let us consider general situations in which several internal degrees of freedom coexist, such as polarization, temporal modes, or orbital angular momentum. A standard way of modeling such systems is to assume that the internal state of each photon factorizes across these degrees of freedom. If $L$ mutually independent internal degrees of freedom are present, the internal state of the photon entering the $i$th spatial mode can be written as
\begin{equation}
\ket{\phi_i}=\bigotimes_{l=1}^L |\phi_{i}^{(l)}\rangle,
\end{equation}
where $|\phi_i^{(l)}\rangle$ denotes the component of the internal state associated with the $l$th degree of freedom. The corresponding internal distinguishability matrix has entries $S^{\mathrm{int}}_{i,j} =\prod_{l=1}^{L}\langle\phi^{(l)}_i|\phi^{(l)}_j\rangle$. 
Thus, additional internal degrees of freedom can only reduce or preserve the pairwise indistinguishability of the photons. In this setting, the distinguishability matrix can be expressed as the Hadamard product of the distinguishability matrices associated with each degree of freedom,
\begin{equation}
\label{eq:severals_S_int}
        S^{\mathrm{int}}= \bigodot_{l=1}^{L} S^{(l)},
\end{equation}
and the multimode bunching probability takes the form
\begin{equation}
    P_\kappa=\text{perm} \Big(H \odot \bigodot_{l=1}^{L} S^{(l)}\Big).
\end{equation}

\subsection{Multimode Bunching Conjecture}
\label{sec:The_standard_indistinguishability_intuition}

Boson bunching phenomena are a direct manifestation of bosonic interference, arising from the symmetrization of the wave function for indistinguishable bosons. Since introducing distinguishability  reduces interference, physical intuition suggests that the bunching probability can only decrease when photons are made more distinguishable. To investigate this expected direct link between indistinguishability and boson bunching, we first consider the standard case of single-mode boson bunching ($|\kappa|=1$), in which one focuses on events when all photons exit through a single output mode $k$. Note that referring to these events as ``single-mode bunching'' is an abuse of terminology, since the photons are actually bunching in a single spatial mode but potentially multiple internal modes that are not resolved by the detector. We will nevertheless use this terminology throughout the paper for simplicity as it is standard in the literature. In this context, Eq.~\eqref{eq:multimode_boson_bunching} reduces to
\begin{equation}
\label{eq:single-mode_boson_bunching}
P_{k}=P_k^{\mathrm{cl}}\,\perm{S^{\mathrm{int}}}.
\end{equation}
Here, $P_k^{\mathrm{cl}}$ corresponds to the classical bunching probability $\prod^n_{i=1}|U_{k,i}|^2$. We define the \textit{quantum enhancement} observed in this bunching event as the ratio between the quantum and classical bunching probabilities. In the present single-mode case, this ratio is equal to $\perm{S^{\mathrm{int}}}$, thereby capturing the increase in bunching probability due to bosonic interference \cite{tichy2015_partial_distinguishability}. The quantum enhancement ranges from $1$ for completely distinguishable bosons to $n!$ for fully indistinguishable ones. 
Thus, $\perm{S^{\mathrm{int}}}$ provides a quantitative measure of indistinguishability \cite{shchesnovich2015tight,tichy2015_partial_distinguishability, shchesnovich2016universality}, with $\perm{S^{\mathrm{int}}}/n!$ reflecting the weight of the permutationally symmetric component of the state $\ket{\Phi}=\ket{\phi_1}\ket{\phi_2}\dots\ket{\phi_n}$. In the single-mode case, it is straightforward to see that the bunching probability $P_{k}$ increases monotonically with this measure of indistinguishability and so it is maximized when the photons are fully indistinguishable (it suffices to consider pure input states when maximizing $P_{k}$ since the latter is linear in the state $\ket{\Phi}\bra{\Phi}$). Therefore, it is natural to ask whether a similar property extends to multimode bunching. Precisely, the following conjecture can be formulated. 

\begin{conjecturep}[Multimode Bunching Conjecture]
\label{conjp:generalizedBunching}
Consider an arbitrary linear interferometer $U$, an arbitrary separable input state of $n$ classically correlated photons, and any subset $\kappa$ of output modes. Then the probability that all output photons bunch into $\kappa$ attains its \emph{global} maximum when the photons are perfectly indistinguishable. 
\end{conjecturep} 
Refs.~\cite{shchesnovich2016universality,shchesnovich2016permanent} provided substantial support for this hypothesis, which was explicitly formulated as a conjecture in Ref.~\cite{bosonbunching}. Conjecture~P\ref{conjp:generalizedBunching} admits a direct mathematical translation: for arbitrary matrices $H$ and $S^{\mathrm{int}}$, one has
\begin{equation}
\label{eq:conjectureP1}
\perm{H \odot S^{\mathrm{int}}} \leq \perm{H}.
\end{equation}
When all photons are indistinguishable, the internal-state matrix $S^{\mathrm{int}}$ coincides with the matrix of all ones $\mathbb{E}$; hence, $\perm{H \odot \mathbb{E}}=\perm{H}$. 
As noted in Ref.~\cite{shchesnovich2016universality}, this conjecture is closely connected to a longstanding mathematical conjecture on the permanent of Hadamard products due to Bapat and Sunder.

\begin{conjecturem}[Bapat--Sunder 1985 \cite{bapat1985majorization}]
\label{conjm:bapatSunder_1}
Let $A$ and $B$ be $n\times n$ positive semidefinite Hermitian matrices. Then  
\begin{equation}
\label{eq:conj1}
    \Perm\,(A\odot B) \leq \Perm\,(A) \prod_{i=1}^n B_{ii}.
\end{equation}
\end{conjecturem} 

Thus, Conjecture~P\ref{conjp:generalizedBunching} would hold provided Conjecture~M\ref{conjm:bapatSunder_1} is true, implying that multimode bunching would then behave, qualitatively, just as single-mode bunching.

\subsection{Anomalous boson bunching}

A first indication that the relation between indistinguishability and bunching becomes more subtle in the multimode regime was provided in Refs.~\cite{shchesnovich2016universality,shchesnovich2016permanent}. These works showed that, for a specific interferometer, a five-boson state with entangled internal degrees of freedom can bunch more than the corresponding fully indistinguishable state. This effect is related to another conjecture on matrix permanents, namely the Permanent-On-Top (POT) conjecture~\cite{Soules1966}, but it does not contradict Conjecture~P\ref{conjp:generalizedBunching} because the input state is no longer required to be separable (an entangled state is needed to beat the indistinguishable state). In other words,
the validity of Conjecture~P\ref{conjp:generalizedBunching}, which is the natural multimode extension of HOM,
 had remained unresolved.

\begin{figure*}[t]
  \centering
  \subfloat[
  \label{fig:2a}]{
    \includegraphics[width=0.34\textwidth]{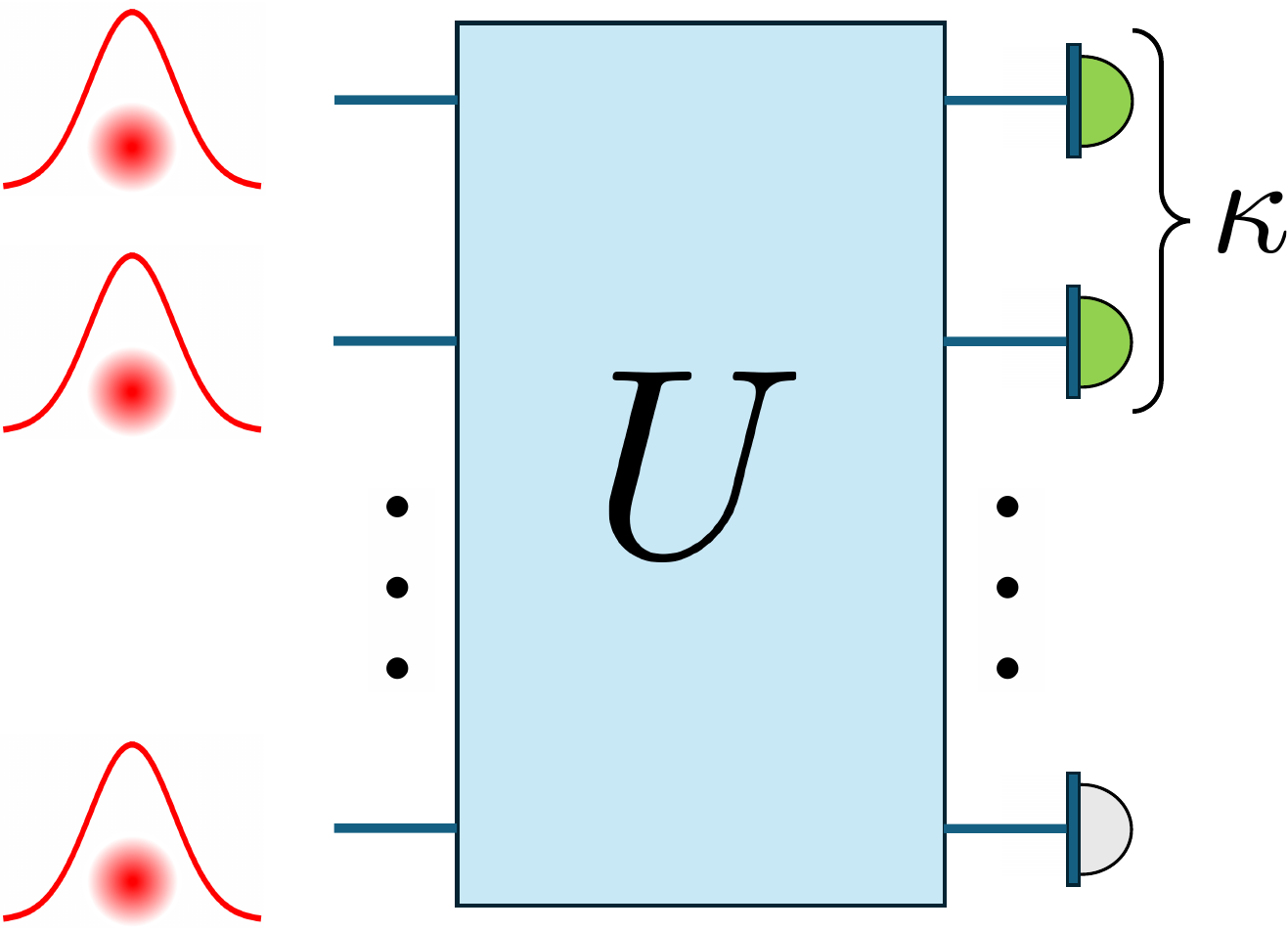}
  }
  \hspace{0.14\textwidth}
  \subfloat[
  \label{fig:2b}]{
    \includegraphics[width=0.34\textwidth]{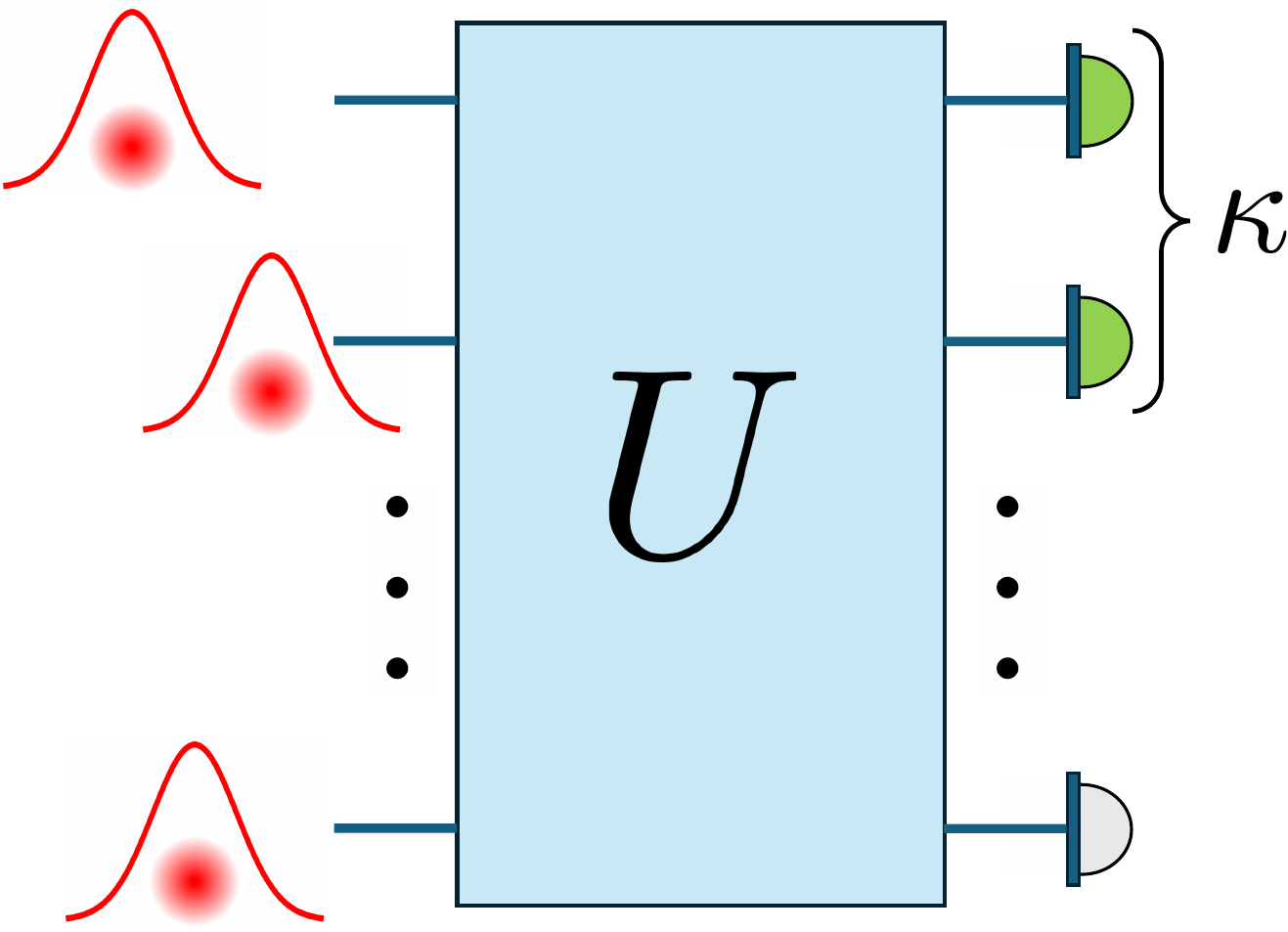}
  }
  \caption{ \justifying 
  Illustration of the anomalous bunching phenomenon through a comparison of two scenarios.
$\bm{(}\mathbf{a}\bm{)}$ In the first scenario, the photons have identical internal states and arrive simultaneously at the interferometer.
$\bm{(}\mathbf{b}\bm{)}$ In the second scenario, the interferometer is unchanged, but the photons are partially distinguishable in an internal degree of freedom, here represented by different time delays. This temporal mismatch is only a concrete example; the same mechanism may arise from other internal degrees of freedom.
Since partial distinguishability typically reduces interference, one might expect the bunching probability in the second scenario to be lower than in the first one. However, by exploiting counterexamples to Conjecture~M\ref{conjm:bapatSunder_1}, the opposite behavior can occur, leading to anomalous bunching, as demonstrated in Refs.~\cite{bosonbunching,pioge2023anomalous,pioge2026validation}.
  }
  \label{fig:standard_anomalous}
\end{figure*}
The next decisive development came in 2016, more than thirty years after the formulation of the Bapat--Sunder conjecture, when Drury constructed two explicit $7\times 7$ matrices disproving Conjecture~M\ref{conjm:bapatSunder_1}~\cite{drury2016}. This mathematical counterexample was subsequently exploited in Ref.~\cite{bosonbunching} to engineer physical counterexamples to Conjecture~P\ref{conjp:generalizedBunching}. In particular, the authors exhibited an interferometric configuration involving  a seven-photon product state in which partial distinguishability enhances the two-mode bunching probability beyond the value attained in the fully indistinguishable regime. 
Following the terminology of Ref.~\cite{pioge2023anomalous}, we refer to configurations in which a separable, partially distinguishable input state leads to a larger bunching probability than the fully indistinguishable bosonic state as \textit{anomalous bunching}. An instance of this effect, where photon distinguishability arises from time delays, is illustrated in Fig.~\ref{fig:standard_anomalous}.

A closer analysis of the counterexamples found in Ref.~\cite{bosonbunching} showed that the states exhibiting anomalous bunching deviate substantially from the fully indistinguishable bosonic state. This observation, supported by additional arguments, raised the question of whether nearly indistinguishable bosons may possibly also undergo anomalous bunching. In Ref.~\cite{pioge2023anomalous}, the authors answered in the affirmative and presented an eight-photon state that is arbitrarily close to the fully indistinguishable bosonic state and exhibits a slightly higher bunching probability than the fully indistinguishable state.

Motivated by the goal of using the multimode bunching probability as a certification criterion for a boson sampler, the authors of Refs.~\cite{geller2025,pioge2026validation} identified a few families of interferometric configurations and partial distinguishability scenarios in which anomalous bunching is ruled out. Furthermore, since uncontrolled time delays in photon arrival times constitute a major source of distinguishability in current photonic platforms, the question of whether such a source of distinguishability can lead to anomalous bunching was particularly important. In Ref.~\cite{pioge2026validation}, the authors resolved this question by presenting a 16-photon scenario in which time-delay-induced distinguishability alone enhances the multimode bunching probability.

\section{TOTAL INDISTINGUISHABILITY}
\label{sec:New_interpretation_of_the_bunching_probability}
The various counterexamples to Conjecture~P\ref{conjp:generalizedBunching} call for a reassessment of the connection between boson bunching and  indistinguishability. In particular, they suggest that the distinguishability of the input photons in terms of internal degrees of freedom does not give the full picture. As we argue below, a proper picture should also incorporate the distinguishability induced by postselection at the output of the interferometer through the photons’ spatial degrees of freedom. The resulting framework is then based on what we coin the \textit{total indistinguishability}.

To see this, let us examine the (sub-normalized) postselected single-photon states obtained by restricting the output state to the subset $\kappa$. The output state associated with the photon injected into the $i$th input mode is
\begin{equation}
\ket{\psi^{\kappa}_i}=\ket{s_i^\kappa}\ket{\phi_i},
\end{equation}
where 
\begin{equation}
\ket{s_i^\kappa}=\sum_{k\in \kappa} U_{k,i}\ket{k},
\end{equation}
describes the sub-normalized spatial state of the photon (within the subset $\kappa$ of output modes) after propagation through the interferometer. We now consider the inner product between two such postselected  single-photon output states,
\begin{align}
    \braket{\psi^{\kappa}_i}{\psi^{\kappa}_j}&=\braket{s_i^\kappa}{s_j^\kappa}\braket{\phi_i}{\phi_j} \nonumber\\
    &=\sum_{k\in \kappa} U_{k,i}^{*} U_{k,j}\braket{\phi_i}{\phi_j}\nonumber\\
&=H_{i,j}\, S^{\mathrm{int}}_{i,j}. 
\end{align}
The identity $H_{i,j}=\braket{s_i^\kappa}{s_j^\kappa}$ offers a clear physical interpretation of the matrix $H$ that was defined in Eq.~\eqref{eq:definition-matrix-H}. It can be interpreted as an unnormalized distinguishability matrix associated with the spatial degrees of freedom restricted to the output modes in $\kappa$. Since the states $\ket{s_i^\kappa}$ are not normalized, the diagonal elements of $H$ are not necessarily equal to one, in contrast to those of $S^{\mathrm{int}}$. This is corrected by defining the distinguishability matrix $S^{\kappa}$, which is the normalized version of $H$, namely
\begin{equation}
    S^{\kappa}_{i,j}=\frac{\braket{s_i^\kappa}{s_j^\kappa}}{\sqrt{\braket{s_i^\kappa}{s_i^\kappa}\braket{s_j^\kappa}{s_j^\kappa}}}=\frac{H_{i,j}}{\sqrt{H_{i,i}H_{j,j}}},~~\forall i,j, 
\end{equation}
with the convention that $S^{\kappa}_{i,j}=0$ whenever either $\ket{s_i^\kappa}$ or $\ket{s_j^\kappa}$ has zero norm.
By applying a standard property of the permanent to Eq.~\eqref{eq:multimode_boson_bunching}—namely, that a common multiplicative factor can be pulled out from any row or column, here, the normalization factors $\sqrt{H_{i,i}}$ and $\sqrt{H_{j,j}}$—the multimode bunching probability takes the compact form 
\begin{equation}
\label{eq:compact-form}
    P_{\kappa}=\prod^n_{i=1} H_{i,i}~\perm{S^{\kappa}\odot S^{\mathrm{int}}}.
\end{equation}
It is worth noting that $\prod^n_{i=1} H_{i,i}$ simply corresponds to the bunching probability of classical (or fully distinguishable) particles in $\kappa$, which we denote as $P_\kappa^{\mathrm{cl}}$. Indeed, when the photons are distinguishable,  $S^{\mathrm{int}}$ is the identity matrix, so that $\perm{S^{\kappa} \odot S^{\mathrm{int}}}=1$, and consequently $P_{\kappa} = \prod_{i=1}^{n} H_{i,i}$. Thus, as before, the permanent in Eq.~\eqref{eq:compact-form} captures the \textit{quantum enhancement} of the bunching probability. The Hadamard-product structure of $S^{\kappa}\odot S^{\mathrm{int}}$ mirrors Eq.~\eqref{eq:severals_S_int}, indicating that spatial distinguishability within $\kappa$ can be treated simply as one degree of freedom among others. This motivates the definition of the \textit{total distinguishability matrix} 
\begin{equation}
S^{\mathrm{tot}}=S^{\kappa}\odot S^{\mathrm{int}},
\end{equation}
which combines the internal and spatial contributions to distinguishability within $\kappa$. Furthermore, the definition of $S^{\mathrm{tot}}$ allows the multimode bunching probability to be written in a unified form, 
\begin{equation}
\label{eq:proba_bunching_general}
    P_{\kappa}= P_\kappa^{\mathrm{cl}} \,\perm{S^{\mathrm{tot}}}.
\end{equation}
This last equation generalizes the standard formula for single-mode boson bunching, Eq.~\eqref{eq:single-mode_boson_bunching}. Indeed, if $\kappa$ contains only one output mode, then $S^{\kappa}=\mathbb{E}$,  and Eq.~\eqref{eq:single-mode_boson_bunching} is recovered. The normalized quantity $\perm{S^{\mathrm{tot}}}/n!$ is then equal to the weight of the permutationally symmetric component of the normalized product state
\begin{equation}
|\widetilde{\Psi}^{\kappa}\rangle
=
|\tilde{\psi}^{\kappa}_1\rangle
|\tilde{\psi}^{\kappa}_2\rangle
\cdots
|\tilde{\psi}^{\kappa}_n\rangle,
\end{equation}
where the states $|\tilde{\psi}^{\kappa}_i\rangle$ are the normalized versions of the postselected one-photon states $\ket{\psi^{\kappa}_i}$.  Accordingly, it serves as a measure of \textit{total indistinguishability} within $\kappa$, accounting for all degrees of freedom, both internal and spatial. 

Building on this notion of total indistinguishability, we are now in a position to reformulate Conjecture~P\ref{conjp:generalizedBunching} into a provable statement, given in the following proposition.
\newlength{\facetcornerwidth}

\begin{table*}[t]
\centering
\renewcommand{\arraystretch}{1}
\setlength{\tabcolsep}{8pt}
\setlength{\facetcornerwidth}{\dimexpr0.13\textwidth+2\tabcolsep\relax}
\small
\begin{tabular}{
|>{\centering\arraybackslash}m{0.13\textwidth}
|>{\centering\arraybackslash}m{0.36\textwidth}
|>{\centering\arraybackslash}m{0.39\textwidth}|}
\hline

\multicolumn{1}{|@{}c@{}|}{%
\begin{tikzpicture}[baseline=(current bounding box.center)]
  \path[use as bounding box] (0,0) rectangle (\facetcornerwidth,0.9cm);
  \draw[line width=0.35pt] (0,0.9cm) -- (\facetcornerwidth,0);
  \node at (0.22\facetcornerwidth,0.25cm) {$S^{A}$};
  \node at (0.72\facetcornerwidth,0.65cm) {$S^{B}$};
\end{tikzpicture}%
}
&
\facetheadercell{Internal\\distinguishability}
&
\facetheadercell{Spatial\\distinguishability}
\\
\hline

\facetbodycell{Spatial\\distinguishability}
&
\facetbodycell{
\textit{Multimode bunching with}\\
\textit{partially distinguishable photons.}\\
\hyperref[sec:standard_anomalous_bunching]{Sec.~\ref*{sec:standard_anomalous_bunching}}
}
&
\facetbodycell{
\textit{Multimode bunching with indistinguishable}\\
\textit{photons in spatial superpositions of input modes.}\\
\hyperref[sec:several_interferometers]{Sec.~\ref*{sec:several_interferometers}}
}
\\
\hline

\facetbodycell{Internal\\distinguishability}
&
\facetbodycell{
\textit{Single-mode bunching with two independent}\\
\textit{internal degrees of freedom.}\\
\hyperref[sec:Extra_distinguishability]{Sec.~\ref*{sec:Extra_distinguishability}}
}
&
\facetbodycell{
\textit{Quantum enhancement of multimode}\\
\textit{bunching vs single-mode bunching.}\\
\hyperref[sec:Multimode_outperform_single-mode]{Sec.~\ref*{sec:Multimode_outperform_single-mode}}
}
\\
\hline

\end{tabular}

\caption{
Classification of the four anomalous bunching facets discussed in Sec.~\ref{sec:new_facets}.
Rows and columns specify the physical origin of the distinguishability matrices $S^A$ and $S^B$ in Eq.~\eqref{eq:general_anomalous}, leading to four anomalous bunching scenarios.
}
\label{tab:all_facets}
\end{table*}

\begin{proposition}[Maximal multimode bunching follows from highest total indistinguishability]
\label{proposition:generalizedBunching}
Consider an arbitrary linear interferometer $U$, an arbitrary separable input state of $n$ classically correlated photons, and any subset $\kappa$ of output modes. Then the probability that all photons bunch into $\kappa$ attains its global maximum when the total indistinguishability of the photons within $\kappa$ is maximal.
\end{proposition}
In other words, the direct link between multimode bunching and total indistinguishability is recovered and straightforwardly extends the case of single-mode bunching. Proposition~\ref{proposition:generalizedBunching} also implies that, for a fixed interferometer $U$ and subset $\kappa$ (hence a fixed $S^{\kappa}$), the input distinguishability matrix $S^{\mathrm{int}}$ that maximizes the bunching probability is the one that maximizes the total indistinguishability, i.e., $\perm{S^{\kappa}\odot S^{\mathrm{int}}}$. Since the Bapat--Sunder conjecture can be violated, this maximum can be realized by an internal distinguishability matrix  $S^{\mathrm{int}} \ne \mathbb{E}$ for some well-chosen spatial distinguishability matrices $S^{\kappa}$. Hence, Proposition~\ref{proposition:generalizedBunching} does not preclude anomalous bunching while preserving the intuitive link between bunching and total indistinguishability. Furthermore,  it uncovers several previously unexplored facets of anomalous bunching, as presented in Sec.~\ref{sec:new_facets}.

\section{Other facets of anomalous boson bunching}
\label{sec:new_facets}

Before presenting the different facets of anomalous bunching, let us first explain why a violation of the Bapat--Sunder conjecture seems counterintuitive from a mathematical viewpoint. Consider two distinguishability matrices $S^{A}$ and $S^{B}$. Since the modulus of every entry of a distinguishability matrix lies between $0$ and $1$, their Hadamard product satisfies $|(S^{A}\odot S^{B})_{i,j}|\leq |S^{A}_{i,j}|$,  $\forall i,j$. One might therefore expect $\perm{S^{A}\odot S^{B}}$ to be maximized by choosing $S^{B}=\mathbb{E}$, since the permanent is expressed as a sum of products of $(S^{A}\odot S^{B})_{i,j}$ entries. The Hadamard product, however, does not merely reduce the modulus of the entries; it also alters their phases. In some cases, by appropriately choosing $S^{B}_{i,j}$, one can modify the phases in $(S^{A}\odot S^{B})_{i,j}$  so that the sum entering the permanent increases, while the modulus of each term decreases,
thereby leading to a violation of Conjecture~M\ref{conjm:bapatSunder_1}. An alternative interpretation of this violation in terms of the fully symmetric component of an $n$ bipartite separable quantum state is provided in Appendix~\ref{appendix:generalized_indistinguishabilities}.

Returning to multimode boson bunching, this effect shows up in the following way. Consider partially distinguishable photons described by a distinguishability matrix $S^{A}$. Adding an independent degree of freedom described by $S^{B}$ can only reduce the pairwise indistinguishability of the photons. Nevertheless, the resulting modification of their collective phases can increase $\perm{S^{A}\odot S^{B}}$, and hence their $n$-photon indistinguishability. The distinguishability matrices $S^{A}$ and $S^{B}$ may have different physical origins, as each can be associated with either internal or spatial degrees of freedom of the photonic wave functions. Consequently, the condition that signals an anomaly, namely
\begin{equation}
\label{eq:general_anomalous}
    \perm{S^{A} \odot S^{B}}>\perm{S^{A}},
\end{equation}
admits different physical realizations, depending on whether $S^{A}$ and $S^{B}$ originate from an internal or spatial degree of freedom. The four possible cases are summarized in Table~\ref{tab:all_facets} and lead to four different facets of anomalous bunching. Sec.~\ref{sec:standard_anomalous_bunching} corresponds to the original anomalous bunching scenario, while the other three cases give rise to additional manifestations of this anomaly, as discussed below.

\begin{figure*}[t!]
  \centering
  \subfloat[Polarization distinguishability\label{fig:3a}]{
    \includegraphics[width=0.48\textwidth]{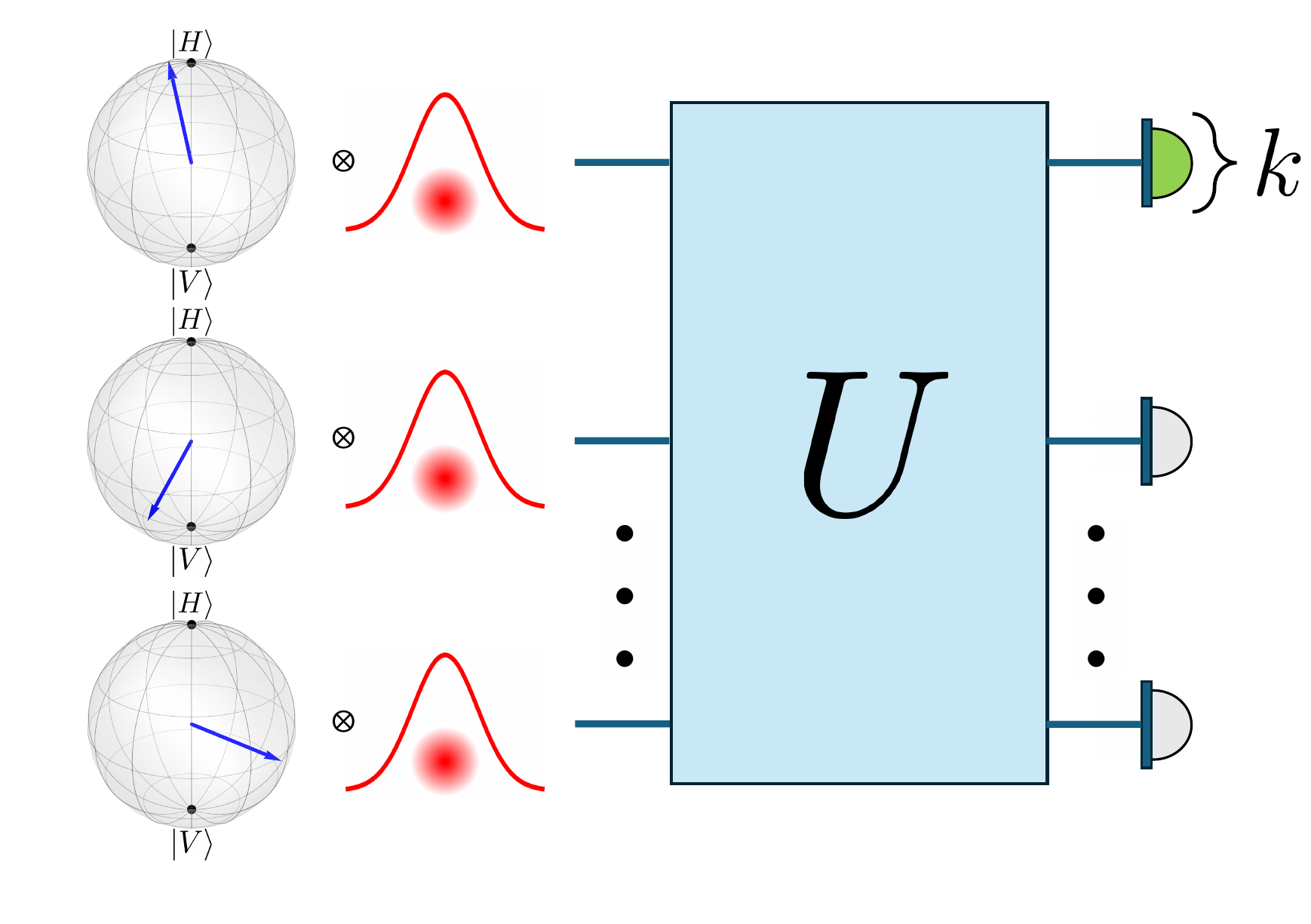}
  }
  \hfill
  \subfloat[Polarization and time delay  distinguishability\label{fig:3b}]{
    \includegraphics[width=0.48\textwidth]{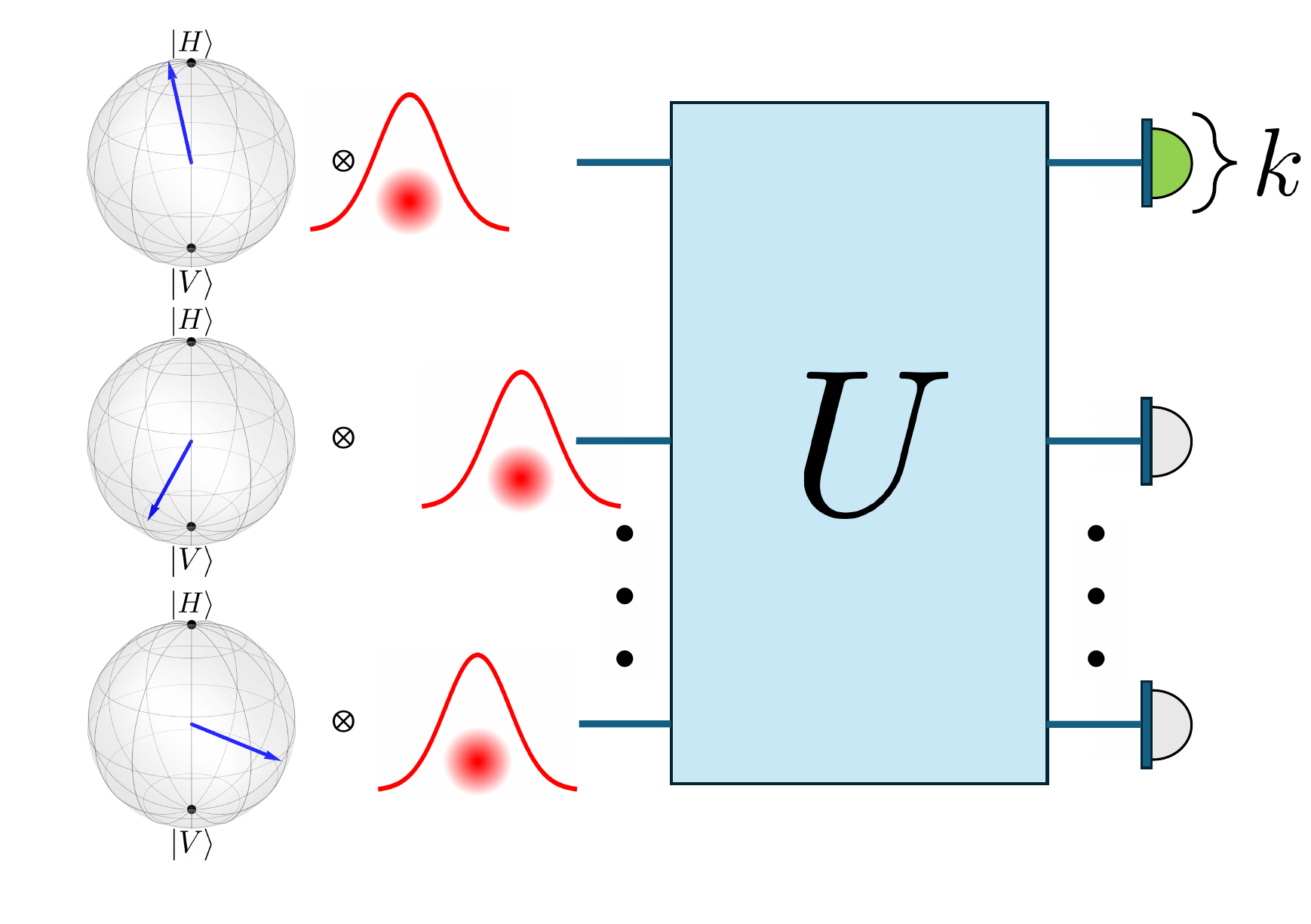}
  }
  \caption{ \justifying 
  Interferometric configurations illustrating the facet of anomalous bunching discussed in Sec.~\ref{sec:Extra_distinguishability} through the comparison of two scenarios. $\bm{(}\mathbf{a}\bm{)}$ The first configuration corresponds to single-mode photon bunching in the first output mode. The photons arrive simultaneously at the interferometer but differ in polarization. Distinguishability therefore arises solely from this degree of freedom. $\bm{(}\mathbf{b}\bm{)}$ In the second configuration, the interferometer and the polarization states remain unchanged, but the photons additionally differ in their arrival times. One might expect that introducing an additional internal degree of freedom can only reduce single-mode bunching, since it reduces the pairwise indistinguishability of the photons. However, by exploiting counterexamples to Conjecture~M\ref{conjm:bapatSunder_1}, the opposite behavior can occur: introducing specific time delays between the photons can increase the single-mode bunching probability. This effect can arise for any interferometer and any choice of the single spatial output mode $k$.  This reveals a new manifestation of anomalous bunching, for which an explicit counterexample is constructed in Appendix~\ref{appendix:Concrete_extra_dist}.}
    \label{fig:extra_dist}
\end{figure*}
\subsection{Partially distinguishable particles can enhance multimode bunching beyond the case of  indistinguishable particles}
\label{sec:standard_anomalous_bunching}
The interpretation based on total indistinguishability as introduced in Sec.~\ref{sec:New_interpretation_of_the_bunching_probability} allows us to revisit the phenomenon of anomalous bunching discussed in \cite{bosonbunching, pioge2023anomalous, pioge2026validation}. This effect is revealed by comparing two situations. In the first, the photons are indistinguishable, and we consider the bunching probability in the subset $\kappa$ of output modes, given by $P_{\kappa} = P_\kappa^{\mathrm{cl}}\,\perm{S^{\kappa}}$.  In the second, the interferometer and subset $\kappa$ remain the same, while the photons exhibit partial internal distinguishability, yielding a bunching probability $P_{\kappa}=P_\kappa^{\mathrm{cl}}\,\perm{S^{\kappa}\odot S^{\mathrm{int}}}$. Since the classical bunching probabilities $P_\kappa^{\mathrm{cl}}$ are identical in both cases, anomalous bunching arises when partially distinguishable photons yield a larger quantum enhancement than fully indistinguishable ones, that is, when
\begin{equation}
\label{eq:anomalous_nat_phot}
    \perm{S^{\kappa} \odot S^{\mathrm{int}}}>\perm{S^{\kappa}}.
\end{equation}
Such configurations contradict Conjecture~P\ref{conjp:generalizedBunching}, but they are fully consistent with Proposition~\ref{proposition:generalizedBunching}. Due to the interplay between spatial and internal distinguishability, the total indistinguishability $\perm{S^{\mathrm{tot}}}$ can be larger in the second case than in the first, since there exist counterexamples to Conjecture~M\ref{conjm:bapatSunder_1}, with $A$ and $B$ taken as Gram matrices. Yet, Proposition~\ref{proposition:generalizedBunching} reestablishes the direct connection between indistinguishability and boson bunching.

\subsection{Additional internal distinguishability can enhance single-mode bunching}
\label{sec:Extra_distinguishability}

\begin{figure*}[t!]
  \centering
  \subfloat[Single-mode boson bunching \label{fig:4a}]{
    \includegraphics[width=0.34\textwidth]{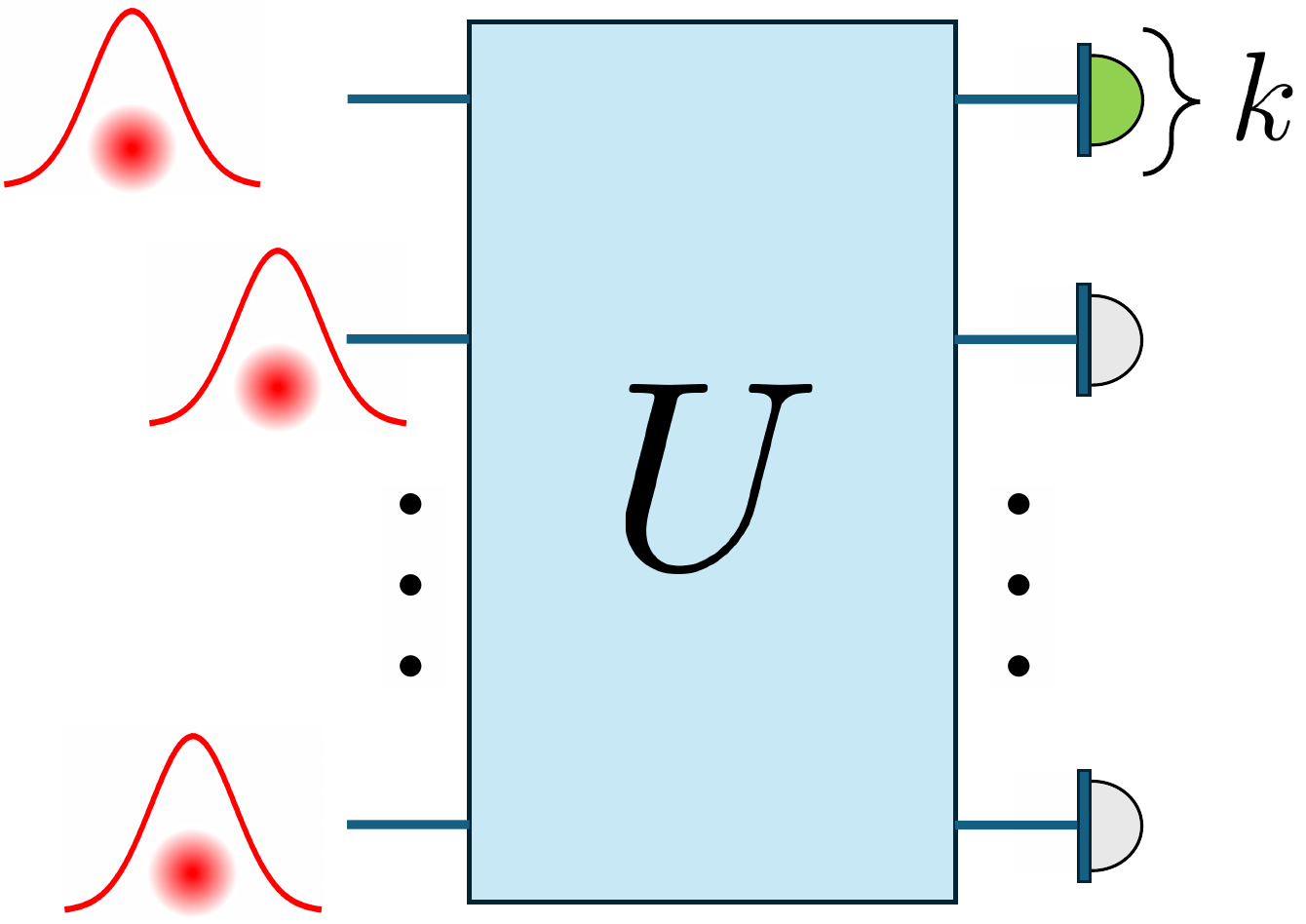}
  }
  \hspace{0.14\textwidth}
  \subfloat[Multimode boson bunching \label{fig:4b}]{
    \includegraphics[width=0.34\textwidth]{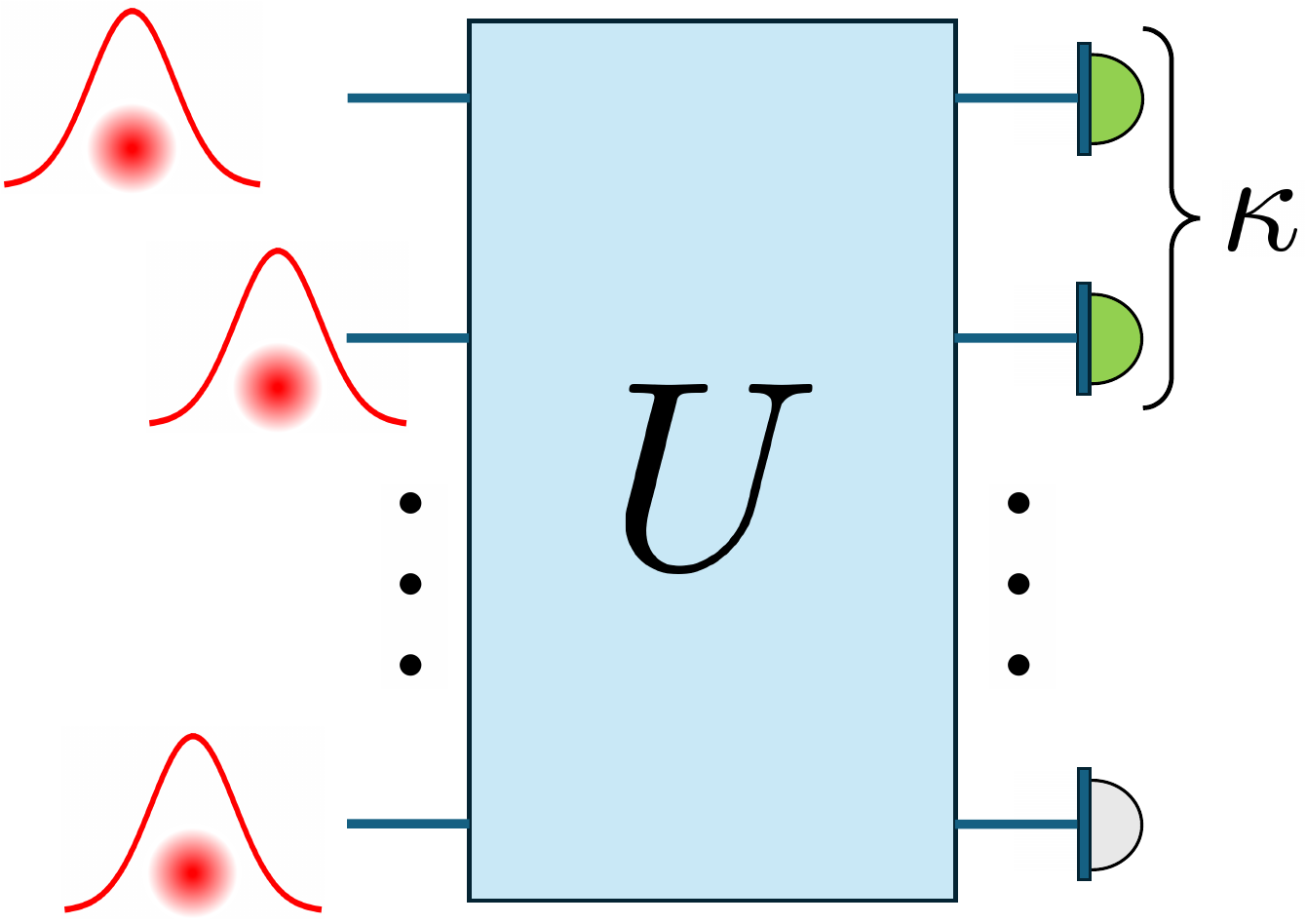}
  }
  \caption{ \justifying 
  Comparison of two interferometric configurations illustrating the facet of anomalous bunching discussed in Sec.~\ref{sec:Multimode_outperform_single-mode}. In contrast to the previous facets, the comparison here concerns the quantum enhancement factor, i.e., the ratio of the quantum bunching probability to its classical counterpart, rather than the bunching probability itself. $\bm{(}\mathbf{a}\bm{)}$ The first configuration corresponds to single-mode boson bunching in a given output mode $k$, with partial internal distinguishability, here illustrated by time delays between the photons. $\bm{(}\mathbf{b}\bm{)}$ In the second scenario, the time delays between the photons remain unchanged, but we consider multimode boson bunching into a subset $\kappa$ of output modes. In this case, the photons can be distributed among different spatial modes within $\kappa$, and the quantum enhancement factor is affected by both internal and spatial degrees of freedom. One might expect single-mode bunching to provide the largest possible quantum enhancement. However, by exploiting counterexamples to Conjecture~M\ref{conjm:bapatSunder_1}, the opposite behavior can occur: multimode bunching can exhibit a larger quantum enhancement factor than single-mode bunching with partially distinguishable particles. An explicit example of this phenomenon is presented in Appendix~\ref{appendix:output_patterns}. 
  }
\label{fig:multi_outperform_single}
\end{figure*}
In its original form, the anomalous bunching effect can arise only for multimode bunching. Here, we identify a distinct counterintuitive phenomenon that occurs instead in the single-mode bunching situation. Remarkably, this effect is independent of the interferometer. It is again revealed by comparing two physically relevant scenarios.
In the first scenario, we analyze the bunching probability at a single output mode $k$, and the internal state of each photon is characterized by a single degree of freedom (e.g., polarization). In this case, we have that $S^{\kappa} = \mathbb{E}$ and the internal state of the $i$th photon is thus $\ket{\phi_i} = \ket{\phi_i^{A}}$, which defines the distinguishability matrix $S^{A}$ with entries $S^{A}_{i,j} = \braket{\phi_i^{A}}{\phi_j^{A}}$. In the second scenario, we again focus on single-mode boson bunching in mode $k$, but introduce an additional internal degree of freedom (e.g., the arrival time) while keeping the first unchanged. The internal state of the $i$th photon is then
\begin{equation}
\ket{\phi_i} = \ket{\phi_i^{A}} \otimes \ket{\phi_i^{B}},
\end{equation}
which leads to the following distinguishability matrix
\begin{equation}
S^{\mathrm{tot}}=S^{A}\odot S^{B},
\end{equation}
where $S^{A}_{i,j}=\braket{\phi^{A}_i}{\phi^{A}_j}$ and $S^{B}_{i,j}=\braket{\phi^{B}_i}{\phi^{B}_j}$. 
By analogy with the HOM effect, one might expect that adding an independent source of distinguishability between the photons, thereby reducing their pairwise indistinguishability, can only effect a decrease in the quantum-enhanced bunching probability.
However, it is possible to contradict this intuition by exploiting counterexamples to Conjecture~M\ref{conjm:bapatSunder_1}. This anomalous bunching effect  corresponds to the special case of Eq.~\eqref{eq:general_anomalous} where both distinguishability matrices arise from internal degrees of freedom.  Contrary to the original case (Sec.~\ref{sec:standard_anomalous_bunching}), this manifestation of anomalous bunching can occur for single-mode bunching in any linear interferometer. 

Fig.~\ref{fig:extra_dist} provides an example of a physical setting in which the introduction of a time delay between the photons can surprisingly increase the single-mode bunching probability. This could in principle be realized via the counterexample from \cite{pioge2026validation}, which focuses on Gram matrices realizable by time delays; however, the example found would require a 16-photon experiment.  In Appendix~\ref{appendix:Concrete_extra_dist}, we present a simpler example based on the independent manipulation of polarization and time-bin degrees of freedom, and we provide the explicit internal state preparations of seven photons that would lead to this effect. We also compute the values of the resulting bunching probabilities using the Fourier interferometer, since this interferometer maximizes the magnitude of this effect. The observation of the effect would require precise estimations of single-mode bunching probabilities of the order of $10^{-4}$, which remains  challenging for current technology.

On a related note, it has already been experimentally demonstrated that decreasing the pairwise distinguishability between the input photons can increase the single-mode bunching probability, even in three-photon scenarios \cite{rodari2024}. However, this was achieved by tuning collective photonic phases through the joint manipulation of the temporal and polarization degrees of freedom. This scenario differs from what we consider here, since the connection to the Bapat--Sunder inequality arises only when the two internal degrees of freedom are tuned independently, preserving the tensor-product structure of the internal wave function of the input single-photon states. 

\subsection{Multimode bunching can exhibit a larger quantum enhancement than single-mode bunching}
\label{sec:Multimode_outperform_single-mode}

\begin{figure*}[t]
  \centering
  \subfloat[
  \label{fig:5a}]{
    \includegraphics[width=0.4\textwidth]{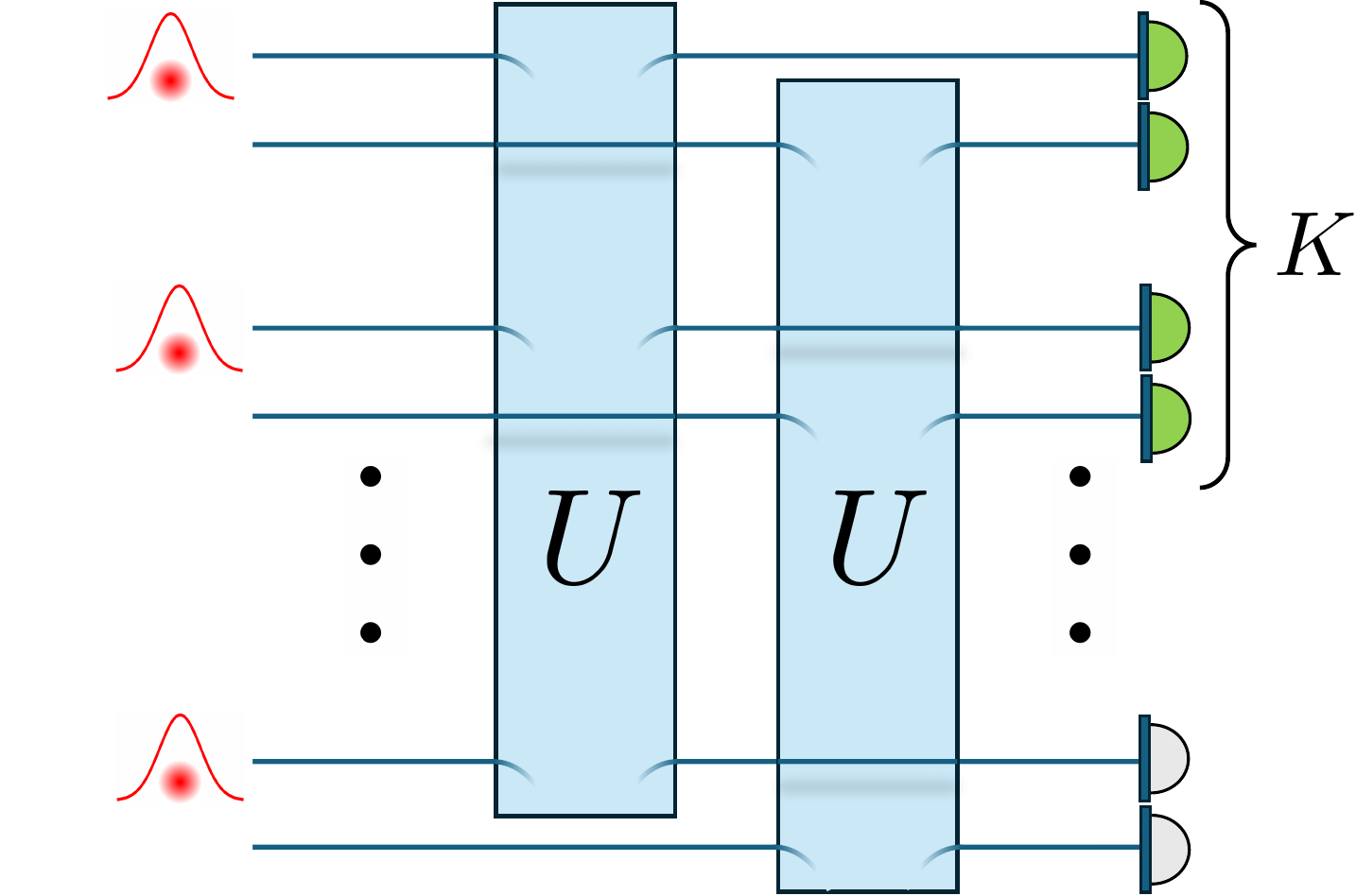}
  }
  \hspace{0.14\textwidth}
  \subfloat[
  \label{fig:5b}]{
    \includegraphics[width=0.43\textwidth]{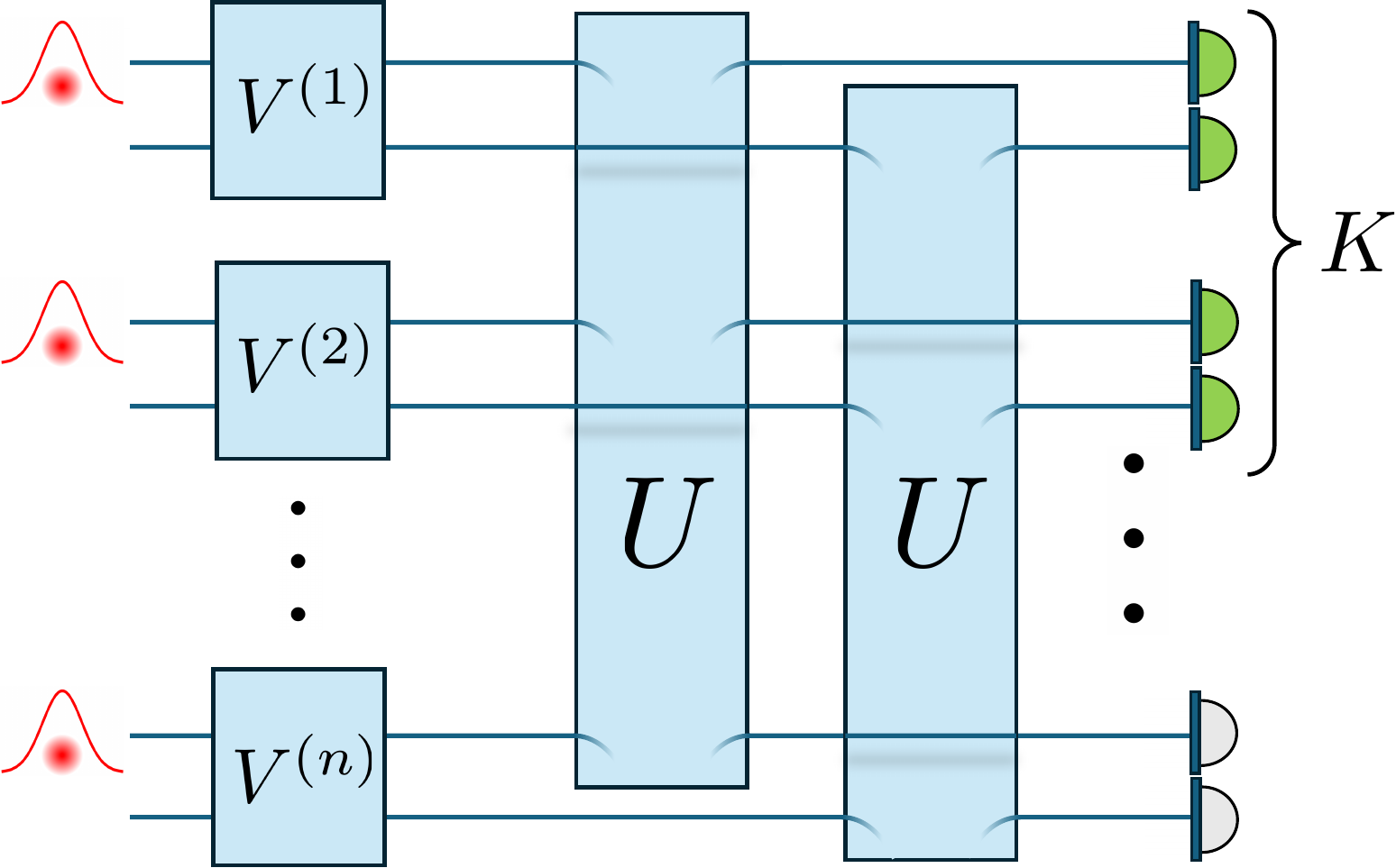}
  }
  \caption{ \justifying 
  Comparison of two interferometric configurations illustrating the facet of anomalous bunching discussed in Sec.~\ref{sec:several_interferometers}. $\bm{(}\mathbf{a}\bm{)}$ In the first scenario, indistinguishable bosons enter the interferometer $U$. We consider the multimode bunching probability within the subset $K$. $\bm{(}\mathbf{b}\bm{)}$ In the second scenario, the photons still occupy the same internal mode but possess an additional spatial degree of freedom. Each photon first enters a distinct interferometer acting on this additional spatial degree of freedom, and then passes through a second layer of identical interferometers $U$, each implementing the same transformation as in the first scenario. One might expect the bunching probability within $K$ to be lower in the second scenario, since the photons are spread across several interferometers. However, by exploiting counterexamples to Conjecture~M\ref{conjm:bapatSunder_1}, the opposite behavior can occur: preparing specific superpositions over the different interferometers can enhance the multimode bunching probability.
  }
\label{fig:severals_interf}
\end{figure*}

Single-mode boson bunching is often regarded as one of the clearest manifestations of bosonic interference and has therefore motivated extensive investigations \cite{tichy2015_partial_distinguishability,rodari2024,Spagnolo_General_Rules,Niu:09}. This interest is further justified since, for fully indistinguishable particles, the quantum enhancement attains $n!$, its maximum value. In a multimode bunching event, certain output configurations involve photons occupying different spatial modes within $\kappa$. As a result, these configurations are less spatially concentrated than single-mode bunching events, in which all photons occupy the same output mode. Furthermore, once the spatial distribution of the photons within $\kappa$ differs, the maximal quantum enhancement of $n!$ can no longer be achieved. Consequently, it is natural to ask whether the quantum enhancement of any bunching event involving partially distinguishable bosons is always upper-bounded by $\perm{S^{\mathrm{int}}}$, that is, the quantum enhancement obtained in the single-mode bunching regime.

However, the following example disproves this statement, giving a physical meaning to the inequality 
\begin{equation}
\label{eq:anomalous_Qenhancement}
    \perm{ S^{\mathrm{int}} \odot S^{\kappa}}>\perm{S^{\mathrm{int}}}.
\end{equation}
We remark that, compared to Eq.~\eqref{eq:anomalous_nat_phot}, the roles of the spatial and internal distinguishability matrices are interchanged.  Indeed, the  left-hand side of Eq.~\eqref{eq:anomalous_Qenhancement} is the quantum enhancement of the multimode bunching probabilities for input photons with internal distinguishability matrix $S^{\mathrm{int}}$, whereas the right-hand side of Eq.~\eqref{eq:anomalous_Qenhancement} is the quantum enhancement of any single-mode bunching probability with the same input state.

Fig.~\ref{fig:multi_outperform_single} illustrates such a configuration, in which a multimode bunching event exhibits a higher quantum enhancement than single-mode bunching with partially distinguishable particles. As the quantum enhancement of multimode bunching can exceed $\perm{S^{\mathrm{int}}}$, it is natural to examine the quantum enhancement factor associated with each output pattern in which all photons occupy modes in $\kappa$. In view of Eq.~\eqref{eq:anomalous_Qenhancement}, it is straightforward to see that certain output patterns within this family exhibit a larger quantum enhancement factor than single-mode bunching. A detailed discussion of a concrete counterexample is provided in Appendix~\ref{appendix:output_patterns}.  We remark that, since the matrices $S^{\kappa}$ and $  S^{\mathrm{int}}$ 
need to be carefully constructed to realize a counterexample to the Bapat--Sunder conjecture, in most multiphoton interference experiments involving photons with differences in their internal degrees of freedom,  the probability of single-mode bunching events will be more significantly enhanced than multimode bunching events, when compared to the behavior of classical particles (distinguishable photons).

\subsection{Anomalous bunching in nested spatial interferometers}
\label{sec:several_interferometers}
Let us move to the physical interpretation of anomalous bunching when  the relevant distinguishability matrices in the underlying Bapat--Sunder inequality arise solely from spatial degrees of freedom. This last facet can be seen as a realization of the original setting where anomalous bunching was found (Sec.~\ref{sec:standard_anomalous_bunching}) but without relying on internal distinguishability. Instead, additional interferometers generate spatial superpositions that reproduce the role played by the internal states in the original setting. The scenario is thus the following. We consider $d$ copies of the same $m\times m$ interferometer $U$, that is, the action of the interferometer in the single-particle space of $m\times d$ spatial modes is given by $\hat{U}\otimes \mathbb{\hat{1}}_d$. It is natural to label the single-particle states as $\ket{i, j}$, representing the $i$th spatial mode of the $j$th copy of the interferometer $U$.  Fig.~\ref{fig:severals_interf} represents the case $d=2$ for simplicity. We consider now two different configurations: either the photons are all sent to the first interferometer $U$ (one in each of the first $n$ modes), each one in state $\ket{i, 1}$, or they are first sent through additional interferometers $V^{(i)}$, which create superpositions of input modes of the form $\ket{i,\phi_i}$ before the photons enter the $d$ copies of $U$. For $d=2$, this superposition can be prepared by beam splitters as shown in Fig.~\ref{fig:severals_interf}. The subset of output modes we consider for multimode bunching is $K= \{\ket{i, j}, i\in \kappa,\, j\in \{1,\ldots,d\}\}$. By construction, the equation describing multimode bunching in this scenario is identical to the one in the original scenario of   Sec.~\ref{sec:standard_anomalous_bunching}, and
the bunching probability can be expressed as
\begin{equation}
P_{K}=P^{\mathrm{cl}}_{\kappa}\,\perm{S^{\kappa}\odot S^{v}}.
\end{equation}
However, while $S^{\kappa}$ is the same distinguishability matrix as in the first scenario, $S^{v}$ now encodes the spatial distinguishability induced by the additional layer of linear interferometers $V^{(i)}$. It would be natural to expect that the multimode bunching in $K$ would be maximized if all photons interfere together in the same interferometer $U$, reaching the bunching probability 
\begin{equation}\label{eq:joint_interference}
P_{\kappa}=P^{\mathrm{cl}}_{\kappa}\,\perm{S^{\kappa}},
\end{equation}
since the action of the additional interferometers $V^{(i)}$ decreases the spatial pairwise indistinguishability within $K$.
In fact, it can be seen that if the $i$th photon were sent probabilistically to one of the copies of $U$, i.e., were sent to mode $\ket{i,j}$ with probability $p^{(i)}_j$, the bunching probability would indeed always be smaller than the value given by Eq.~\eqref{eq:joint_interference}. However, by spreading the photons over the different copies in specific spatial superposition states, it is possible to realize a counterexample to Conjecture~M\ref{conjm:bapatSunder_1} and achieve higher multimode bunching.

\section{CONCLUSION}

Motivated by the apparent tension between the discovered existence of anomalous bunching and the common intuition linking boson bunching to indistinguishability, we introduced a physical framework that resolves this discrepancy (see Proposition~\ref{proposition:generalizedBunching}). The central notion of this framework is the total indistinguishability of the postselected output state, which combines internal overlaps with the spatial overlaps induced by the interferometer and the restriction to the set $\kappa$ of measured modes. This viewpoint restores a direct connection between bunching and total indistinguishability via Eq.~\eqref{eq:proba_bunching_general}, while naturally accommodating anomalous bunching. This anomaly arises precisely in configurations in which reducing the pairwise internal indistinguishability of the photons can nevertheless increase their total indistinguishability, hence their bunching probability. 

Building on this framework, we uncovered three additional manifestations of anomalous bunching, depending on the chosen physical realization of the matrices $A$ and $B$ appearing in Conjecture~M\ref{conjm:bapatSunder_1}. We found that $(i)$~adding an independent internal source of distinguishability can enhance multimode and even single-mode bunching; $(ii)$~multimode bunching events can exhibit a larger quantum enhancement than single-mode bunching events with partially distinguishable photons; and $(iii)$~purely spatial distinguishability in multi-layer interferometric settings can enhance the bunching probability despite reducing pairwise spatial overlaps.
Although these three facets defy physical intuition, they all follow from violations of Conjecture~M\ref{conjm:bapatSunder_1}.

An important remaining challenge is to derive transparent, intuitive criteria for predicting when a reduction in pairwise indistinguishability can lead to an increase in total indistinguishability, and hence to anomalous bunching. While recent progress has identified broad classes of configurations in which the effect is excluded or can be engineered \cite{geller2025,pioge2026validation}, a simple and general characterization of the existence of anomalous bunching remains elusive.

\section*{Acknowledgments}
L.P. is a FRIA grantee of the
Fonds de la Recherche Scientifique -- FNRS and also acknowledges support from the Fonds David et Alice Van Buuren and the Fondation Jaumotte-Demoulin. N.J.C. acknowledges support from the Fonds de la Recherche Scientifique (F.R.S.-FNRS) under Grant No. T.0060.26 as well as under project CHEQS within the Excellence of Science (EOS) program. L.N.  acknowledges funding from FCT-Fundação para a Ciência e a Tecnologia (Portugal) via the Project No. CEECINST/00062/2018. L.N. also acknowledges support from the project with the reference n.º 2023.15565.PEX, funded by national funds through FCT – Fundação para a Ciência e a Tecnologia, I.P., and from Horizon Europe project EPIQUE (Grant No. 101135288).

\bibliographystyle{unsrt}
\bibliography{main}

\onecolumngrid

\appendix

\section{Derivation of the multimode bunching probability in second quantization}
\label{appendix:second_quantizationcalculation}

The multimode bunching probability $P_{\kappa}$ can be computed by projecting the output state 
\begin{equation}
    \ket{\Psi^{\mathrm{out}}} = \hat{U} \prod_{j=1}^n \ad_{j, \phi_j} \ket{0} = \prod_{j=1}^n \left( \sum_{k=1}^m U_{k,j} \, \ad_{k, \phi_j} \right)\ket{0},
\end{equation}
onto the subspace with photons occupying only the output modes in $\kappa$, i.e., the state obtained at the interferometer output by postselecting on the vacuum in every output mode not in $\kappa$. This yields the (unnormalized) state 
\begin{equation}
    \ket{\Psi^{\kappa}} = \prod_{j=1}^n \hat{A}_j^\dagger \ket{0},
\end{equation}
where we have defined the operators 
\begin{equation}
\label{eq:def-operators_A_j}
    \hat{A}_j^\dagger =  \sum_{k\in\kappa} U_{k,j} \, \ad_{k, \phi_j} , \qquad \forall j.
\end{equation} 
Note that the sum over the output spatial modes $k$ is restricted to $\kappa$. The multimode bunching probability can be expressed as the squared norm of $\ket{\Psi^{\kappa}}$, namely
\begin{equation}
\label{eq:squared_norm}
P_\kappa= \bra{0} \hat{A}_n \cdots \hat{A}_1 \hat{A}_1^\dagger \cdots \hat{A}_n^\dagger \ket{0} .
\end{equation}
In order to calculate $P_{\kappa}$, we seek the commutation relations obeyed by the operators $\hat{A}_i$ and $\hat{A}_i^\dagger$. By expressing the internal states in some orthonormal basis $\{\ket{u}\}$, that is,
$\ket{\phi_j}=\sum_u c_{u,j} \ket{u}$ with $c_{u,j}  =  \langle u | \phi_j \rangle$, we may write the creation operators as
$\ad_{k, \phi_j} = \sum_u  c_{u,j} \, \ad_{k,u}$, where $\ad_{k,u}$ creates a photon in the $k$th spatial mode and internal state $\ket{u}$. With the usual commutation relations $[\hat{a}_{l,v},\ad_{k,u}]=\delta_{k,l}\, \delta_{u,v}$ and 
$[\hat{a}_{l,v},\hat{a}_{k,u}] =[\ad_{l,v},\ad_{k,u}]=0$, we obtain
\begin{align}
[\hat{a}_{l,\phi_i},\ad_{k,\phi_j}]&=\delta_{k,l}\, S^{\mathrm{int}}_{i,j}, \nonumber \\ [\hat{a}_{l,\phi_i},\hat{a}_{k,\phi_j}]&=[\ad_{l,\phi_i},\ad_{k,\phi_j}]=0,
\end{align}
where we have defined
\begin{equation}
\label{eq_def_Gram_matrix_ij}
S^{\mathrm{int}}_{i,j}=\braket{\phi_i}{\phi_j}, \qquad i,j\in  \{1,\ldots,n\}.
\end{equation}
The matrix $S^{\mathrm{int}}$, which is referred to as the \emph{distinguishability matrix} \cite{tichy2015_partial_distinguishability,shchesnovich2015partial}, is thus the $n \times n$ Gram matrix constructed with all possible overlaps of the internal states of the photons.
Then, using Eq.~\eqref{eq:def-operators_A_j} and defining 
the $n \times n$ positive semidefinite Hermitian matrix $H$ via
\begin{equation}
\label{eq:H_Shchesnovich}
H_{i,j} = \sum_{k \in \kappa} U_{k,i}^* U_{k,j}, \qquad \forall i,j,
\end{equation}
we obtain the following commutation relations
\begin{align}
\label{eq:commutation_relations}
[\hat{A}_i,\hat{A}_j^\dagger] = \bar{H}_{i,j}, \qquad
[\hat{A}_i,\hat{A}_j] = [\hat{A}^\dagger_i,\hat{A}^\dagger_j] = 0,
\end{align}
where $\bar{H}= H \odot S^{\mathrm{int}}$. Now, it is easy to evaluate  Eq.~\eqref{eq:squared_norm} by exploiting Eqs.~\eqref{eq:commutation_relations}, as well as the fact that $\hat{A}_i \ket{0}=0$. The idea is to move each $\hat{A}_i$ step by step to the right in Eq.~\eqref{eq:squared_norm}, using the commutator each time it crosses an $\hat{A}_j^\dagger$, until it reaches the rightmost position where it annihilates the vacuum $\ket{0}$.

Let us begin with the simplest case, $n=1$. We have
\begin{equation}
P_1= \bra{0} \hat{A}_1 \hat{A}_1^\dagger  \ket{0} = \bra{0} (\bar H_{1,1}+ \hat{A}_1^\dagger \hat{A}_1 )\ket{0} = \bar H_{1,1} ,
\end{equation}
where we have used the commutation relations and $\hat{A}_1 \ket{0} =0$. This is, of course, the permanent of the $1\times 1$ matrix obtained by keeping only the first row and first column of $\bar H $. Similarly, it is trivial to show that 
\begin{equation}
\label{eq-recurrence-n=1}
\bra{0} \hat{A}_i \hat{A}_j^\dagger  \ket{0}  = \bar H_{i,j} ,
\end{equation}
which is the permanent of the $1\times 1$ matrix obtained by keeping the $i$th  row and $j$th column of $\bar H $.

For $n>1$, we need to move the operator $\hat{A}_1$ step by step to the right, exploiting the commutator each time it crosses an $\hat{A}_j^\dagger$ operator. The first move gives
\begin{equation}
 \bra{0} \hat{A}_n \cdots \hat{A}_2 \!\!\!\!\underbrace{\stackrel{\big\downarrow}{\hat{A}_1} \hat{A}_1^\dagger}_{\bar H_{1,1}+\hat{A}_1^\dagger \hat{A}_1}\!\!\!\! \hat{A}_2^\dagger \cdots \hat{A}_n^\dagger \ket{0}  =\bar H_{1,1} \bra{0} \hat{A}_n \cdots \hat{A}_2 \hat{A}_2^\dagger \cdots \hat{A}_n^\dagger \ket{0} + \bra{0} \hat{A}_n \cdots  \hat{A}_2 \hat{A}_1^\dagger \stackrel{\big\downarrow}{\hat{A}_1} \hat{A}_2^\dagger \cdots \hat{A}_n^\dagger \ket{0},
 \label{eq:first-move}
\end{equation}
where we identify the location of the operator $\hat{A}_1$ that is moved with an arrow. The next move consists of transforming the second term in the right-hand side of Eq.~\eqref{eq:first-move} by moving $\hat{A}_1$ again one step to the right, namely  
\begin{equation}
  \bra{0} \hat{A}_n \cdots  \hat{A}_2 \hat{A}_1^\dagger \!\!\!\!\underbrace{\stackrel{\big\downarrow}{\hat{A}_1} \hat{A}_2^\dagger}_{\bar H_{1,2}+\hat{A}_2^\dagger \hat{A}_1}\!\!\!\! \hat{A}_3^\dagger\cdots \hat{A}_n^\dagger \ket{0} = \bar H_{1,2} \bra{0} \hat{A}_n \cdots  \hat{A}_2 \hat{A}_1^\dagger  \hat{A}_3^\dagger\cdots \hat{A}_n^\dagger \ket{0}+ \bra{0} \hat{A}_n \cdots  \hat{A}_2 \hat{A}_1^\dagger  \hat{A}_2^\dagger \stackrel{\big\downarrow}{\hat{A}_1} \hat{A}_3^\dagger\cdots \hat{A}_n^\dagger \ket{0}.
\end{equation}
By recursively applying this same transformation until $\hat{A}_1$ reaches the rightmost position, in which case
\begin{equation}
 \bra{0} \hat{A}_n \cdots  \hat{A}_2 \hat{A}_1^\dagger  \hat{A}_2^\dagger  \hat{A}_3^\dagger\cdots \hat{A}_n^\dagger \stackrel{\big\downarrow}{\hat{A}_1} \ket{0} = 0,
\end{equation}
and putting together all terms of the recurrence, we obtain 
\begin{equation}
 \bra{0} \hat{A}_{n \cdots 1} \hat{A}_{1\cdots n}^\dagger  \ket{0} = \bar H_{1,1} \bra{0} \hat{A}_{n\cdots 2}  \hat{A}_{2\cdots n}^\dagger  \ket{0} + \bar H_{1,2} \bra{0} \hat{A}_{n\cdots 2}  \hat{A}_{13\cdots n}^\dagger   \ket{0} + \cdots + \bar H_{1,n} \bra{0} \hat{A}_{n\cdots 2}  \hat{A}_{1\cdots (n-1)}^\dagger \ket{0} .
\end{equation}
In this expression, we have used the self-explanatory shorthand notation $\hat{A}_{i}\hat{A}_{j}\hat{A}_{k}\equiv \hat{A}_{ijk}$ and 
$\hat{A}^\dagger_{i}\hat{A}^\dagger_{j}\hat{A}^\dagger_{k}\equiv \hat{A}^\dagger_{ijk}$ (note that the ordering of the indices does not matter since the $\hat{A}$ operators -- or $\hat{A}^\dagger$ operators -- commute with one another). By explicitly indicating which operator $\hat{A}_i$ or $\hat{A}_j^\dagger$ is skipped in the above expression, we recognize the Laplace expansion formula for the permanent, namely
\begin{equation}
 \underbrace{\bra{0} \hat{A}_{n \cdots 1} \hat{A}_{1\cdots n}^\dagger  \ket{0}}_{\perm{\bar H}} = \bar H_{1,1} \underbrace{\bra{0} \hat{A}_{n\cdots 2 \cancel{1}}  \hat{A}_{\cancel{1}2\cdots n}^\dagger  \ket{0}}_{\perm{\bar H(1,1)}} + \bar H_{1,2} \underbrace{\bra{0} \hat{A}_{n\cdots 2 \cancel{1}}  \hat{A}_{1\cancel{2}3\cdots n}^\dagger   \ket{0}}_{\perm{\bar H(1,2)}} + \cdots + \bar H_{1,n} \underbrace{\bra{0} \hat{A}_{n\cdots 2\cancel{1}}  \hat{A}_{1\cdots (n-1)\cancel{n}}^\dagger \ket{0}}_{\perm{\bar H(1,n)}} ,
 \label{eq-recurrence-n=n+1}
\end{equation}
where $\bar H(i,j)$ denotes the matrix obtained from $\bar H$ by deleting the $i$th row and $j$th column. Since $\bar H = H \odot S^{\mathrm{int}}$, by applying Eq.~\eqref{eq-recurrence-n=1} together with repeated use of Eq.~\eqref{eq-recurrence-n=n+1}, we obtain that the multimode bunching probability is given by
\begin{equation}
    P_{\kappa}=\perm{H \odot S^{\mathrm{int}}}.
\end{equation}

\section{A quantum-state interpretation of Bapat--Sunder violations using $n$ bipartite states}
\label{appendix:generalized_indistinguishabilities}
Let us consider a set of $n$ bipartite separable quantum states $\ket{\chi_i}= \ket{\chi^A_i}\ket{\chi^B_i}$, where the states $\ket{\chi^A_i}$ are fixed and $\ket{\chi^B_i}$ can be chosen at will. The associated distinguishability matrix is $S^{\mathrm{tot}}_{i,j} = \braket{\chi_i}{\chi_j}= \braket{\chi^A_i}{\chi^A_j}\braket{\chi^B_i}{\chi^B_j}$ and we can also define the distinguishability matrices $S^{A}$, $S^{B} $  with $S^{A/B}_{ij} = \langle \chi^{A/B}_i|{\chi^{A/B}_j}\rangle$. Intuitively, it could be expected that in order to maximize the fully symmetric component of the state $\ket{\chi}=\bigotimes_{i=1}^n \ket{\chi_i}$ we should try to ``align`` all the states $\ket{\chi^B_i}$, choosing them to be some reference state $\ket{\chi^B_0}$. Indeed, this strategy maximizes the pairwise similarity between the states, as measured by  $|S^{\mathrm{tot}}_{ij}|=|\braket{\chi_i}{\chi_j}|$. In fact, for a permutation $\sigma$ of the states $\ket{\chi_i}$, this strategy also maximizes the absolute value of all terms
\begin{equation}
    \bra{\chi}\hat{P}_\sigma\ket{\chi}= \prod_{i=1}^{n} \braket{\chi_i}{\chi_{\sigma(i)}}
\end{equation}
appearing in the computation of the fully symmetric component
\begin{equation}\label{eq:symmetric_component}
\frac{\perm{S^\mathrm{tot}}}{n!}=\frac{1}{n!} \sum_{\sigma \in S_n} \bra{\chi}\hat{P}_\sigma\ket{\chi}. 
\end{equation}
However, the quantities  $\bra{\chi}\hat{P}_\sigma\ket{\chi}$, referred to as generalized indistinguishabilities in \cite{annoni2025incoherent}, are complex-valued and so decreasing their absolute value does not always guarantee that the sum appearing  in Eq.~\eqref{eq:symmetric_component} (which is always positive) would decrease. The violation of the Bapat--Sunder conjecture tells us that there are instances of this problem where aligning all the states $\ket{\chi^B_i}$ is not always the best choice to maximize the fully symmetric component of the state $\ket{\chi}$.

\section{Explicit construction of anomalous bunching via added distinguishability}
\label{appendix:Concrete_extra_dist} 
In this appendix, we present a concrete implementation of the facet of anomalous bunching discussed in Sec.~\ref{sec:Extra_distinguishability}. Specifically, we compare two single-mode bunching scenarios. In the first scenario, the internal distinguishability of the photons arises from a single degree of freedom. In the second scenario, an additional source of distinguishability is introduced through a second internal degree of freedom, while the first one is left unchanged (see Fig.~\ref{fig:extra_dist}). Our implementation is based on the simplest known counterexample to Conjecture~M\ref{conjm:bapatSunder_1}, originally introduced in Ref.~\cite{drury2016}, but other physical situations could be constructed based on other counterexamples for nearly indistinguishable photons~\cite{pioge2023anomalous} or time delays~\cite{pioge2026validation}. This mathematical counterexample was translated into a seven-photon interferometric scheme in Ref.~\cite{bosonbunching}, where it served as the basis for the first proposed setup exhibiting the original form of anomalous bunching. To explain how this counterexample can be physically realized in the context of the present facet of anomalous bunching, we proceed as follows. We first specify the internal states of the photons in the first scenario. We then introduce the corresponding internal states in the second scenario. Although this effect is independent of the specific interferometer, it is possible to optimize the interferometer to increase the magnitude of the effect, and so we propose an interferometer that offers the most promising route toward its experimental observation.

In the first scenario, we introduce partial distinguishability by choosing polarization as the distinguishing degree of freedom. Starting from the matrix introduced in Ref.~\cite{drury2016} and following the procedure described in Ref.~\cite{bosonbunching}, we obtain a seven-photon input state. The first five photons are prepared in the polarization states
\begin{equation}
\ket{\phi^p_i} = \frac{1}{\sqrt{2}}\bigl(\ket{H} + \omega^{i}\ket{V}\bigr), \quad i = 1,\dots,5,
\end{equation} 
where $\omega = e^{2 \pi \mathrm{i}/5}$ denotes the fifth root of unity, and $\ket{H}$ and $\ket{V}$ correspond to horizontal and vertical polarization states, respectively. The two remaining photons are prepared in the states $\ket{H}$ and $\ket{V}$. This configuration gives rise to the polarization distinguishability matrix $S^{\mathrm{pol}}$.

In the second scenario, the photons retain the polarization pattern introduced in the first scenario, but additionally exhibit temporal distinguishability. The internal state of the $i$th photon is therefore given by $\ket{\phi_i} = \ket{\phi_i^p} \otimes \ket{\phi_i^t}$. The first five photons are prepared as
\begin{equation}
\ket{\phi_i}=\ket{\phi_i^p}\otimes\frac{1}{\sqrt{2}}\bigl(\ket{t_1}+\omega^{-i}\ket{t_2}\bigr),\quad i=1,\dots,5,
\end{equation}
where $\ket{t_1}$ and $\ket{t_2}$ denote two orthogonal time-bin states. The remaining two photons are prepared in the states $\ket{\phi_6} = \ket{H} \otimes \ket{t_1}$ and $\ket{\phi_7} = \ket{V} \otimes \ket{t_2}$. Owing to the phase factor $\omega^{-i}$, this temporal distinguishability pattern does not arise naturally and must therefore be implemented actively. The temporal states define a distinguishability matrix $S^{t}$ equal to the transpose of the polarization distinguishability matrix $S^{\mathrm{pol}}$. This pair of matrices, originally identified in Ref.~\cite{drury2016}, constitutes a counterexample to Conjecture~M\ref{conjm:bapatSunder_1}, as witnessed by
\begin{equation}
R=\frac{\perm{S^{\mathrm{pol}}\odot S^t}}{\perm{S^{\mathrm{pol}}}}\approx 1.07.
\end{equation}
Since this enhancement is independent of the choice of unitary transformation and of the output mode in which single-mode bunching is considered, we may choose an interferometer that maximizes the classical single-mode bunching probability summed over all spatial modes. This leads to the seven-mode Fourier interferometer, defined by $U_{i,j}=\frac{1}{\sqrt{7}}\omega_7^{(i-1)(j-1)}$, with $\omega_7=e^{2 \pi \mathrm{i}/7}$. In this case, the classical probability of observing single-mode bunching, summed over all seven output modes, satisfies $\sum_{k=1}^{7} P_{k}^{\mathrm{cl}}=\frac{1}{7^6}\approx8.49\times 10^{-6}$. The corresponding bunching probabilities in the first and second scenarios are then
\begin{align}
&\sum_{k=1}^{7} P_{k}^{\mathrm{cl}}\,\perm{S^{\mathrm{pol}}}\approx 3.82\times 10^{-4},\\
&\sum_{k=1}^{7} P_{k}^{\mathrm{cl}}\,\perm{S^{\mathrm{pol}}\odot S^t}\approx 4.11\times 10^{-4}.
\end{align}
The comparison between the two scenarios shows that introducing an additional, independent internal degree of freedom can enhance the bunching probability, thereby revealing a new facet of anomalous bunching.

\section{Output patterns with quantum enhancement exceeding that of single-mode bunching}
\label{appendix:output_patterns}

\begin{figure}[t]
\centering
\includegraphics[width = 0.62\textwidth]{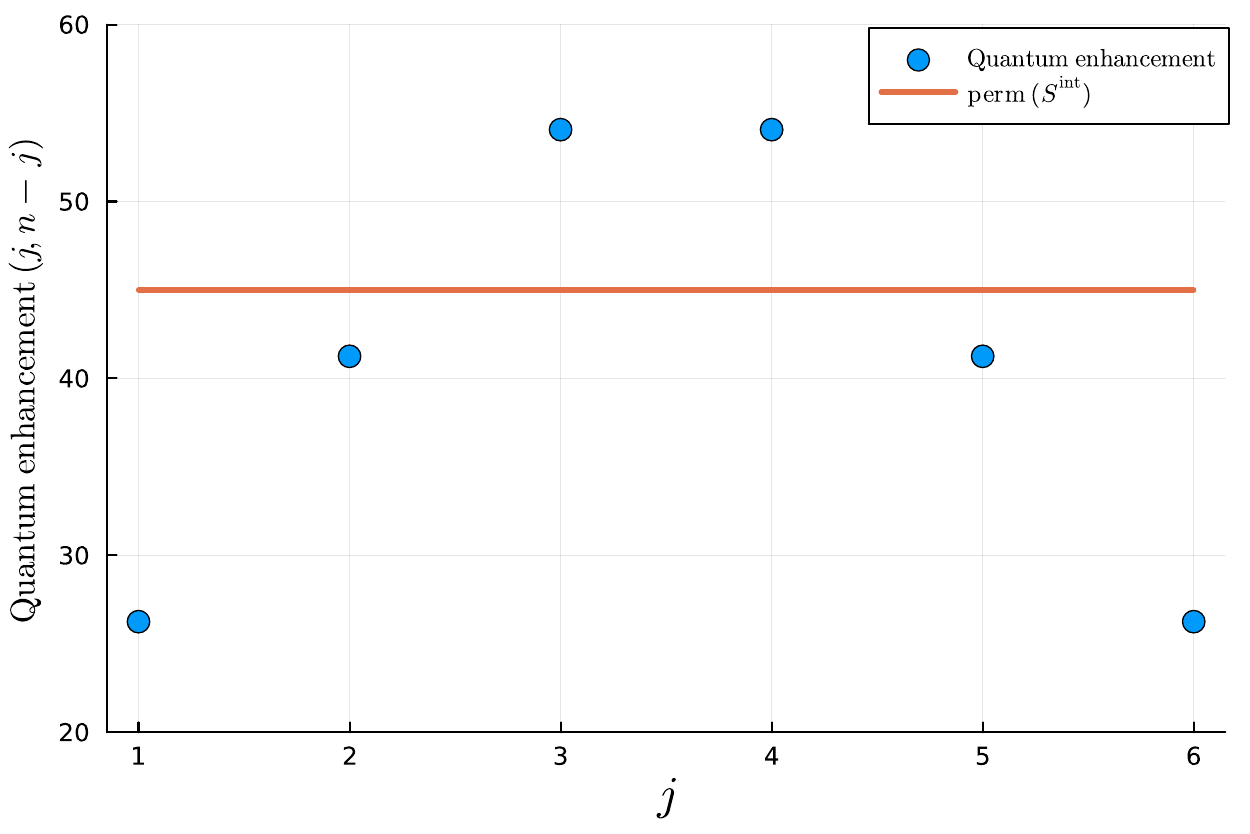}
\caption{\justifying The blue dots represent the quantum enhancement factor associated with each output pattern in which all photons are detected in the two modes of $\kappa$ for the setup described in Appendix~\ref{appendix:output_patterns}. Owing to the symmetry of the circuit, the probabilities of the events $(j,n-j)$ and $(n-j,j)$ are identical. Moreover, the probabilities of the events $(7,0)$ and $(0,7)$ vanish in both the classical and quantum cases and are therefore not shown. The value $\perm{S^{\mathrm{int}}}$, which corresponds to the quantum enhancement of single-mode bunching, is plotted for comparison. The output patterns $(3,4)$ and $(4,3)$ exhibit an enhancement of approximately $54.06$, corresponding to a $20\% $ increase over the single-mode bunching value and indicating anomalous bunching in this configuration.}
\label{fig:Quantum_enhancement_factor}
\end{figure}

For partially distinguishable photons propagating through a linear interferometer, 
we define the quantum enhancement factor of a given output pattern as the ratio between the probability of observing this pattern with partially distinguishable photons and the corresponding probability obtained with classical particles. Sec.~\ref{sec:Multimode_outperform_single-mode} showed that, for partially distinguishable photons, certain output patterns can display a quantum enhancement exceeding that of single-mode boson bunching. In this appendix, we provide an explicit example by analyzing the output patterns of the anomalous bunching configuration introduced in Ref.~\cite{bosonbunching}. This setup involves a seven-photon state whose internal degrees of freedom correspond to the polarization states specified in the first scenario of Appendix~\ref{appendix:Concrete_extra_dist}. The subset $\kappa$ consists of two modes of a seven-mode interferometer, which can be implemented using a five-mode Fourier interferometer followed by two beam splitters. For further details, see Ref.~\cite{bosonbunching}.

The quantum enhancement factor associated with the coarse-grained event in which all photons are detected within $\kappa$, equivalently the classical-probability-weighted average over the corresponding output patterns, is given by $\perm{S^{\kappa}\odot S^{\mathrm{int}}}$. This quantity exceeds $\perm{S^{\kappa}}$ because, by construction, the Gram matrices were chosen so as to violate the Bapat--Sunder conjecture. In the present case, $\perm{S^{\mathrm{int}}}=\perm{S^{\kappa}}$, so no additional modification is required to observe this facet of the phenomenon. In more general situations, such as the one considered in Ref.~\cite{pioge2023anomalous}, observing this effect would require exchanging the roles of $S^{\mathrm{int}}$ and $S^{\kappa}$. Since $\kappa$ contains two spatial modes, the photon-number distribution within $\kappa$ can be expressed using the notation $(j,n-j)$, where $j$ and $n-j$ denote the numbers of photons detected in the first and second output modes of $\kappa$, respectively. The quantum enhancement factors for all output patterns are shown in Fig.~\ref{fig:Quantum_enhancement_factor}. We find that the two output patterns $(3,4)$ and $(4,3)$ exhibit a quantum enhancement factor of approximately $54.06$. Since $\perm{S^{\mathrm{int}}}=45$, these patterns display a quantum enhancement that exceeds that of single-mode bunching by roughly $20.1\%$.

\end{document}